\documentclass[useAMS,usenatbib]{mn2e}
\usepackage{graphicx}
\usepackage{amssymb}
\usepackage{color}
\usepackage{tabularx}
\usepackage{amsmath}

\title[Supernova 2013fc: the big brother of SN 1998S]{Supernova 2013fc in a circumnuclear ring of a luminous infrared galaxy: the big brother of SN 1998S}
\author[Kangas et al.]{T. Kangas$^{1}$\thanks{E-mail:
tjakan@utu.fi}, S. Mattila$^{1,2}$, E. Kankare$^{3}$, P. Lundqvist$^{4}$, P. V\"{a}is\"{a}nen$^{5,6}$, M. Childress$^{7,8}$, \and G. Pignata$^{9,10}$, C. McCully$^{11,12}$, S. Valenti$^{11,12}$, J. Vink\'{o}$^{13,14}$, A. Pastorello$^{15}$, \and N. Elias-Rosa$^{15}$, M. Fraser$^{16}$, A. Gal-Yam$^{17}$, R. Kotak$^{3}$, J.K. Kotilainen$^{2}$, S.J. Smartt$^{3}$, \and L. Galbany$^{10,18}$, J. Harmanen$^{1,19}$, D.A. Howell$^{11,12}$, C. Inserra$^{3}$, G.H. Marion$^{14}$, \and R.M. Quimby$^{20}$, J.M. Silverman$^{14}$, T. Szalai$^{13}$, J.C. Wheeler$^{14}$, C. Ashall$^{21}$, S. Benetti$^{15}$, \and C. Romero-Ca\~nizales$^{10,22}$, K.W. Smith$^{3}$, M. Sullivan$^{23}$, K. Tak\'{a}ts$^{10,9}$ and D.R. Young$^{3}$ \\ \\
(Affiliations can be found after the references)
}

\begin{document}

\pagerange{\pageref{firstpage}--\pageref{lastpage}} \pubyear{2015}

\maketitle

\label{firstpage}

\begin{abstract}

We present photometric and spectroscopic observations of SN 2013fc, a bright type II supernova (SN) in a circumnuclear star-forming ring in the luminous infrared galaxy ESO 154-G010, observed as part of the Public ESO Spectroscopic Survey of Transient Objects (PESSTO). SN 2013fc is both photometrically and spectroscopically similar to the well-studied type IIn SN 1998S and to the bright type II-L SN 1979C. It exhibits an initial linear decline, followed by a short plateau phase and a tail phase with a decline too fast for $^{56}$Co decay with full gamma-ray trapping. Initially the spectrum was blue and featureless. Later on, a strong broad ($\sim 8000$ km s$^{-1}$) H\,$\alpha$ emission profile became prominent. We apply a Starlight stellar population model fit to the SN location (observed when the SN had faded) to estimate a high extinction of $A_V = 2.9 \pm 0.2$ mag and an age of $10_{-2}^{+3}$ Myr for the underlying cluster. We compare the SN to SNe 1998S and 1979C and discuss its possible progenitor star considering the similarities to these events. With a peak brightness of $B = -20.46 \pm 0.21$ mag, SN 2013fc is 0.9 mag brighter than SN 1998S and of comparable brightness to SN 1979C. We suggest that SN 2013fc was consistent with a massive red supergiant (RSG) progenitor. Recent mass loss probably due to a strong RSG wind created the circumstellar matter illuminated through its interaction with the SN ejecta. We also observe a near-infrared excess, possibly due to newly condensed dust.

\end{abstract}

\begin{keywords}
supernovae: general -- supernovae: individual: 2013fc -- galaxies: starburst
\end{keywords}

\section{Introduction}

It is widely accepted that stars more massive than $\sim 8 M_{\odot}$ end their lives in core-collapse supernovae (CCSNe) after evolving through subsequent stages of nuclear burning until they have built up an iron core. The common types of CCSNe are Ib and Ic, with respectively no hydrogen and no hydrogen or helium features in their spectra, and type II with spectral hydrogen features \citep{p1}. Type II supernovae (SNe) in general result from the explosion of a star which has retained at least a part of its hydrogen envelope \citep{p1, p2}, and the differences among the subtypes II-P, II-L and IIb are thought to result from the degree to which the envelope has been retained. The subtype will thus depend on progenitor properties (initial mass, mass loss, metallicity, binarity, rotation). Historically the subtypes II-P and II-L have been thought of as distinct populations of events \citep{p8}, with a long plateau phase of near constant luminosity and a fast linear decline, respectively. \citet{p3} and \citet{faran1,faran2} argued that these subtypes are not continuous, while other studies with larger samples of light curves \citep{p6,p7} have favoured a continuum of properties leading from II-P to II-L, with progenitor mass at zero-age main sequence (ZAMS) likely being the dominant parameter. Using the rise times of type II SNe, \citet{gonzalez} find no evidence of a bimodal II-P/II-L population, while \citet{gall} rule out significantly smaller progenitor radii for II-L. 

The exact, quantitative definition of a type II-L SN is somewhat ambiguous. Spectroscopically, they tend to exhibit strong H\,$\alpha$ emission without the P Cygni absorption profile common for type II-P \citep{p5}. Photometrically, \citet{faran2} classified an event as II-L if its \emph{V}-band magnitude declined more than 0.5 mag from the peak value in the first 50 days after the explosion, while \citet{p9} applied the same criterion in the \emph{R} band; originally, Barbon et al. (1979) defined them as having a linear decline of about 0.05 mag d$^{-1}$ in \emph{R}. Due to the photometric continuum they found between the subtypes, \citet{p6} suggested classifying both simply as type II, with the `plateau'-phase brightness decline rate as one of the main parameters of the SN.

The plateau in a type II-P SN is believed to be the result of an optically thick hydrogen recombination `wave' propagating through a sizeable H-rich envelope, the fairly constant luminosity caused by a balance of ejecta expansion and cooling. Type II-L SNe are usually significantly brighter in peak absolute magnitude than the more common type II-P \citep{p4,p15}, although they are indicated to simply occupy the bright end of a continuum \citep{p6,p7}. The short or non-existent plateaus in the light curves of type II-L SNe \citep[e.g.][]{13by}, and the lack of a P Cygni absorption, are believed to be caused by a relatively low-mass H envelope that simply has a lower column density of absorbing material, and cannot sustain the recombination required for a constant luminosity for long \citep{g71, 1979C2} -- although \citet{nakar} have proposed that the difference is due to $^{56}$Ni contribution and not the ejecta mass. These differences and the relative rarity of type II-L SNe red-- $\sim 7.5$ or $\sim 3$ per cent of all CCSNe according to \citet{p9} and \citet{eldridge} respectively -- suggest a relatively high progenitor mass \citep[a range of 18 -- 23 $M_{\odot}$ is favoured by][]{p9}, stronger mass loss and thus a lower-mass envelope at the time of explosion. A possible progenitor has been identified for SN 2009hd \citep{p10}, who estimate its mass to be $\lesssim 20 M_{\odot}$. \citet{p11} and \citet{09kr1} proposed a direct detection of the progenitor of SN 2009kr as well. However, \citet{09kr2} noted that the source identified as the progenitor of SN 2009kr is affected by contamination from a small cluster. A review by \citet{smartt2} placed the initial mass of the progenitor of SN 2009hd at $15_{-3}^{+3} M_{\odot}$ \citep[consistently with the $\gtrsim15 M_{\odot}$ suggested for type II-L by][]{faran2}, and suggested an upper limit of 18 $M_{\odot}$ on the mass of most CCSN progenitors -- although luminous blue variable (LBV) progenitors of some type IIn SNe can be more massive \citep[e.g.][]{05gl1,05gl2}. The II-P progenitor range has been estimated as $8.5_{-1.5}^{+1}$ to $16.5 \pm 1.5 M_{\odot}$ by \citet{p2}.

There have been indications that interaction with circumstellar matter (CSM) is present in many type II-L SNe. SN 1998S \citep[e.g.][]{1998S1, 1998S2, 1998S3} was spectroscopically classified as a type IIn SN, but showed very II-L-like behavior in both its later spectra and its light curve \citep{1998S1}, with a remarkable similarity to the type II-L SN 1979C \citep[e.g.][]{1979C1, 1979C2}. SN 1979C itself was abnormally bright for a II-L \citep[as pointed out by][]{p15}, with a luminosity comparable to SN 1998S, possibly as a result of CSM interaction. A high mass loss rate for the progenitors of these events is evident from late-time radio observations \citep[e.g.][]{1979C9, 1979C4, 1998S6}. SN 1998S remains one of the brightest and most well-studied SNe of its class, and has been interpreted as having a red supergiant (RSG) progenitor with possibly asymmetric CSM consisting of two separate shells, caused by separate mass loss events \citep{1998S3, 1998S4, mauerhan}; although the asymmetry of the CSM was disputed by \citet{1998S5}. Strong RSG winds were pointed out as a likely cause for the mass loss. The subtype IIn-L was recently suggested for the several known type IIn SNe that, like SN 1998S, show similarity to type II-L \citep{taddia}. The metallicities of the explosion sites of these SNe were found to be similar to those of SNe II-L and II-P but higher than those of long-lasting type IIn and SN impostors, supporting the possibility of massive RSG progenitors for them. \citet{13by} also raised a question of whether all II-L SNe involve CSM interaction, and suggested that if caught early enough, types II-P, II-L and IIn may be spectroscopically similar. Some transitional events, like SN 2007pk \citep{07pk, 07pk2} and PTF11iqb \citep{11iqb}, were suggested to be somewhere between II-P and II(n)-L in terms of progenitor mass and/or mass loss.

In this paper we consider the case of SN 2013fc. We show that this type II SN, originally classified as type IIn due to host galaxy contamination, closely resembles 1998S-like SNe, and is the brightest of these events so far. We will adopt the term `1998S-like' in this paper to mainly refer to SNe 1998S, 1979C and 2008fq, which we will use below as comparison events. These events were all bright type II SNe that photometrically and spectroscopically resemble type II-L more than slowly-declining type IIn or II-P and showed signs of CSM interaction. The structure of this paper is as follows: in Section 2 we summarize the observational data; in Section 3 we determine the extinction toward SN 2013fc and describe its host galaxy; in Sections 4 and 5 we present the photometry and spectroscopy of SN 2013fc, respectively; in Section 6 we discuss the nature and favoured progenitor scenario of the SN; and finally in Section 7 we present our conclusions.

\section{Observations}

\begin{figure*}
\centering
\begin{minipage}{170mm}
\includegraphics[width=\textwidth]{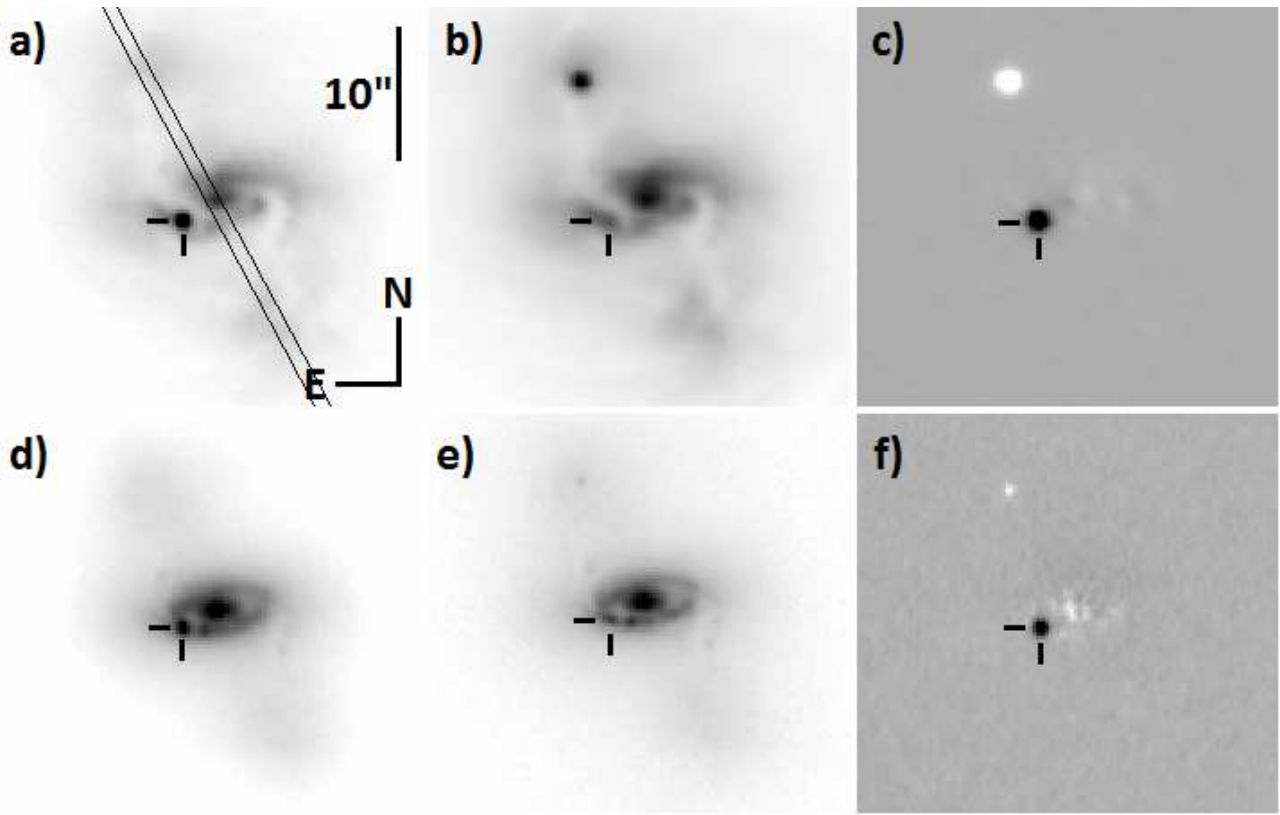}
\caption{The location of SN 2013fc in its host galaxy, ESO 154-G010, marked with tick marks in each image. North is up and east is to the left. a) The NTT/EFOSC2 \emph{V}-band image from 2013 August 29, with the approximate slit position of our August 2014 SALT host galaxy spectrum overplotted. b) The \emph{V}-band template image for NTT/EFOSC2, taken on Nov 16, 2014. The new object in the northern lobe of the galaxy is another supernova, the Type Ia SN 2014eg \citep{atel2, atel3}, serendipitously discovered in these images. SN 2013fc had faded at this time. c) The template-subtracted \emph{V}-band image from 2013 August 29. d) The NTT/SOFI \emph{K}-band image from 2013 October 04, showing the location of the SN coincident with a circumnuclear clumpy ring. e) The \emph{K}-band template for NTT/SOFI, taken on Jan 28, 2015, showing another bright spot in the ring coincident with SN 2013fc, and with SN 2014eg still visible. f) The template-subtracted \emph{K}-band image from 2013 October 04.}
\end{minipage}
\end{figure*}

SN 2013fc was discovered in the galaxy ESO 154-G010 by the CHilean Automatic Supernova sEarch (CHASE; Pignata et al. 2009) on 2013 August 20.2 UT (JD = 2456524.7) using the 41-cm PROMPT1 telescope at Cerro Tololo, Chile \citep{iau1}. The location of the SN in its host galaxy is shown in Figure 1. It was originally classified as a type IIn supernova on 2013 August 29 by \citet{atel1} based on the detection of strong narrow spectral Balmer lines. However, these lines were unresolved in the low-resolution classification spectrum and, based on a sequence of higher-resolution spectra (see Section 3), arose from the host galaxy itself. Thus a better classification of the SN is that of type II-L; see Section 4.

There are no pre-discovery imaging data of ESO 154-G010 closer in time to the discovery of the SN than 2013 July 02.3 UT, 48.9 days before discovery. Therefore determining the explosion date exactly is not possible. However, we can make an estimate based on the similarities to the light curves and spectra of other well studied SNe. The SuperNova IDentification (SNID) package \citep{snid} gives a good match between the spectra of SN 2013fc and the bright type II-L SN 1979C \citep[e.g.][]{1979C5}. SNID also provides a good match between SN 2013fc and SN 1998S \citep[e.g.][]{1998S1,1998S3}, excluding the narrow component of the H\,$\alpha$ line of SN 2013fc, which we show below to originate from the host galaxy. The narrow H\,$\alpha$ emission is only prominent in the SN 1998S spectra taken before the brightness maximum; we also cannot exclude the pre-maximum presence of a narrow H\,$\alpha$ emission line in SN 2013fc itself, since all our spectra were taken after the optical-band maximum. The rise time to maximum for SN 1979C was less than 17 days \citep{1979C1,1979C2}, and for SN 1998S roughly 15 days in the \emph{R} band \citep{1998S2, 98s_early}. For SN 2013fc the \emph{R}-band peak occurred on day 7.2 after its discovery (JD = 2456531.9; see Section 4). Considering the spectroscopic and photometric similarities, we assume the \emph{R}-band rise time to be about 15 days for SN 2013fc as well. We therefore place the shock breakout roughly eight days before the discovery, on August 12.2 (JD = 2456516.7 $\pm$ a few days), and we adopt this as the 0 day epoch in the rest of this paper. All epochs of the photometric and spectroscopic observations reported in this paper are in rest-frame days relative to this explosion date.

Photometric follow-up \emph{UBVRIJHK} observations of SN 2013fc were performed with the 3.6-m New Technology Telescope (NTT) at the European Southern Observatory (ESO), La Silla, Chile, from 2013 August 29 to 2014 March 08. These were done using both the optical instrument ESO Faint Object Spectrograph and Camera v.2 (EFOSC2) and the infrared (IR) instrument Son of ISAAC (SOFI). These observations were carried out as part of the Public ESO Spectroscopic Survey of Transient Objects\footnote{www.pessto.org} \footnote{All reduced PESSTO EFOSC2 and SOFI spectra and imaging data are available from the ESO Science Archive Facility (http://archive.eso.org/cms.html) as PESSTO release number 3, which covers the period between August 2013 and April 2014. Details on how to access the data are available on the PESSTO website.} \citep[PESSTO;][]{pessto2}. Photometric follow-up was also performed in the \emph{BVgri} bands between 2013 August 26 and 2013 December 14 using the Las Cumbres Observatory Global Telescope network \citep[LCOGT;][]{lcogt1}. The LCOGT 1-m telescopes used are located in Siding Spring Observatory (SSO), New South Wales, Australia (1m0-3, 1m0-11); Cerro Tololo, Chile (1m0-4, 1m0-5, 1m0-9); and Sutherland, South Africa (1m0-12, 1m0-13). The Sinistro and SBIG instruments were used. Additional optical photometry was obtained using the PROMPT1 telescope and, later, the 60-cm TRAPPIST telescope in La Silla, from the discovery until November 2013. For the apparent and absolute photometry, see Section 4. In order to subtract the bright background contamination from the photometry, template images were obtained for NTT, PROMPT1 and TRAPPIST in the autumn of 2014. Further sets of NTT/SOFI \emph{JHK} images were obtained in January and September 2015 to ascertain the disappearance of the SN from the earlier templates. Templates for LCOGT were obtained in the autumn of 2015.

Low-resolution optical ($R = 355$ -- 595 depending on grism) and near-infrared (NIR; $R = 550$ -- 611) spectroscopic observations of the SN were obtained with the NTT instruments EFOSC2 and SOFI at La Silla, respectively, as part of the PESSTO survey. Additional low-resolution spectra ($R = 300$) were obtained with the 10-m Southern African Large Telescope (SALT) at the South African Astronomical Observatory, Northern Cape, South Africa, using the Robert Stobie Spectrograph \citep[RSS;][]{rss}. In addition, higher-resolution spectra (R = 3000 in the blue and either 3000 or 7000 in the red part) of the inner region of the host galaxy, including the SN, were obtained with the Wide-field Spectrograph \citep[WiFeS;][]{wifes1,wifes2} integral field unit (IFU) at the Australian National University (ANU) 2.3-m telescope at the Siding Spring Observatory. In total, optical spectra of the SN at fifteen epochs (17 -- 142 days) and one NIR spectrum (82 days) were obtained.

In addition, to better characterize the host galaxy, long-slit spectra were obtained in August 2014 with SALT/RSS, as part of a larger luminous infrared galaxy (LIRG) survey. The slit with PA=24 covered the nucleus and the lobes at the ends of the bar structure (see Figure 1). As such the slit did not cover the SN location but the ring hosting the SN around the nucleus could be characterized. A template spectrum of the location of the SN was obtained with WiFeS in January 2015. From this data cube, a spectrum of the host galaxy core was also extracted. A full list of all our spectroscopic observations, with details of the grisms and gratings used, is presented in Table 1. See Section 5 for the analysis of the spectra. The SN spectra will be available on the Weizmann Interactive Supernova data REPository\footnote{http://www.weizmann.ac.il/astrophysics/wiserep/} \citep{wiserep}.

\begin{table*}
\centering
\begin{minipage}{169mm}
\caption{A log of the spectroscopic observations of SN 2013fc and its host galaxy. Resolutions and exposure times are given for each grism or grating separately, the former including the effect of the slit width. The slit width reported for WiFeS is the width of one IFU slitlet.}
\begin{tabular}[t]{lcccccccc}
    \hline
        JD & Epoch\footnote{Since explosion, in the rest frame of the SN.} & Instrument & Grism/grating & $\lambda$ & Slit & Resolution & Seeing & Exposure time\\
	(2400000+) & (days) & & & (\AA) & ($\arcsec$) & (km s$^{-1}$) & ($\arcsec$) & (s) \\
    \hline
    \hline
	56533.9 & 17 & NTT/EFOSC2 & \#13 & 3650 -- 9250 & 1.0 & 845 & 1.1 & 1500\\
	56539.5 & 22 & SALT/RSS & PG300 & 3500 -- 10000 & 1.25 & 1000 & 1.0 & 1000\\
	56543.7 & 27 & NTT/EFOSC2 & \#11, \#16 & 3345 -- 9995 & 1.0 & 765, 504 & 1.3 & 3600, 3600\\
	56545.2 & 28 & ANU 2.3-m/WiFeS & B3000, R7000 & 3500 -- 7300 & 1.0 & 105, 45 & 2.5 & 1200, 1200\\
	56549.1 & 32 & ANU 2.3-m/WiFeS & B3000, R3000 & 3500 -- 7300 & 1.0 & 105, 105 & 2.5 & 1200, 1200\\
	56555.1 & 38 & ANU 2.3-m/WiFeS & B3000, R3000 & 3500 -- 9800 & 1.0 & 105, 105 & 1.5 & 1200, 1200\\
	56558.4 & 41 & SALT/RSS & PG300 & 3500 -- 10000 & 1.25 & 1000 & 1.0 & 1400\\
	56568.6 & 51 & NTT/EFOSC2 & \#11, \#16 & 3345 -- 9995 & 1.0 & 765, 504 & 0.8 & 3600, 3600\\
	56574.2 & 57 & ANU 2.3-m/WiFeS & B3000, R7000 & 3500 -- 7300 & 1.0 & 105, 45 & 1.7 & 1200, 1200\\
	56590.0 & 72 & ANU 2.3-m/WiFeS & B3000, R7000 & 3500 -- 7300 & 1.0 & 105, 45 & 2.5 & 1200, 1200\\
	56599.7 & 82 & NTT/SOFI & GB, GR & 9500 -- 25200 & 1.0 & 300, 300 & 0.8 & 2160, 2160\\
	56600.8 & 83 & NTT/EFOSC2 & \#11, \#16 & 3345 -- 9995 & 1.0 & 765, 504 & 1.0 & 3600, 3600\\
	56558.4 & 89 & SALT/RSS & PG300 & 3500 -- 10000 & 1.25 & 1000 & 0.8 & 1250\\
	56611.1 & 93 & ANU 2.3-m/WiFeS & B3000, R7000 & 3500 -- 7300 & 1.0 & 105, 45 & 1.7 & 1200, 1200\\
	56628.7 & 110 & NTT/EFOSC2 & \#11 & 3345 -- 7470 & 1.5 & 1148 & 1.9 & 2700\\
	56661.6 & 142 & NTT/EFOSC2 & \#13 & 3650 -- 9250 & 1.0 & 845 & 1.0 & 4500\\
	56901.2 & 378 & SALT/RSS & PG900, PG1800 & 3770 -- 6980 & 1.25 & 270, 95 & 1.6 & 1100, 900\\
	57051.9 & 526 & ANU 2.3-m/WiFeS & B3000, R7000 & 3500 -- 7300 & 1.0 & 105, 45 & 1.5 & 1200, 1200\\
\hline
\end{tabular}
\end{minipage}
\end{table*}

\subsection{Data reduction}

\begin{table*}
\centering
\begin{minipage}{173mm}
\caption{The coordinates (J2000.0) and magnitudes of the local calibration stars used for the optical photometry of SN 2013fc.} 
\begin{tabular}[t]{lccccccc}
    \hline
	Star & RA & Dec & \emph{U} & \emph{B} & \emph{V} & \emph{R} & \emph{I} \\
    \hline
    \hline
	1 & $02^{\mathrm{h}} 45^{\mathrm{m}} 11\fs18$ & $-55\degr 42\arcmin 53\farcs0$ & 22.307 $\pm$ 0.306 & 22.412 $\pm$ 0.064 & 21.145 $\pm$ 0.059 & 20.221 $\pm$ 0.031 & 19.617 $\pm$ 0.025 \\
	2 & $02^{\mathrm{h}} 45^{\mathrm{m}} 04\fs61$ & $-55\degr 43\arcmin 27\farcs9$ & 19.215 $\pm$ 0.031 & 20.153 $\pm$ 0.011 & 19.755 $\pm$ 0.010 & 19.475 $\pm$ 0.011 & 19.186 $\pm$ 0.012 \\
	3 & $02^{\mathrm{h}} 45^{\mathrm{m}} 10\fs40$ & $-55\degr 43\arcmin 57\farcs2$ & 20.506 $\pm$ 0.063 & 19.285 $\pm$ 0.007 & 17.893 $\pm$ 0.007 & 16.938 $\pm$ 0.006 & 16.111 $\pm$ 0.005 \\
	4 & $02^{\mathrm{h}} 45^{\mathrm{m}} 58\fs91$ & $-55\degr 44\arcmin 10\farcs8$ & 16.090 $\pm$ 0.022 & 16.290 $\pm$ 0.004 & 15.835 $\pm$ 0.004 & 15.470 $\pm$ 0.005 & 15.130 $\pm$ 0.007 \\
	5 & $02^{\mathrm{h}} 45^{\mathrm{m}} 15\fs50$ & $-55\degr 45\arcmin 07\farcs1$ & 20.092 $\pm$ 0.045 & 18.912 $\pm$ 0.006 & 17.461 $\pm$ 0.006 & 16.485 $\pm$ 0.006 & 15.655 $\pm$ 0.005 \\
	6 & $02^{\mathrm{h}} 45^{\mathrm{m}} 00\fs59$ & $-55\degr 45\arcmin 30\farcs2$ & 18.286 $\pm$ 0.024 & 17.360 $\pm$ 0.005 & 16.189 $\pm$ 0.004 & 15.411 $\pm$ 0.005 & 14.767 $\pm$ 0.006 \\
	7 & $02^{\mathrm{h}} 45^{\mathrm{m}} 57\fs62$ & $-55\degr 45\arcmin 37\farcs1$ & 21.107 $\pm$ 0.084 & 19.868 $\pm$ 0.008 & 18.429 $\pm$ 0.008 & 17.452 $\pm$ 0.006 & 16.644 $\pm$ 0.005 \\
	8 & $02^{\mathrm{h}} 45^{\mathrm{m}} 20\fs84$ & $-55\degr 46\arcmin 02\farcs6$ & 20.317 $\pm$ 0.054 & 19.740 $\pm$ 0.008 & 18.723 $\pm$ 0.007 & 18.051 $\pm$ 0.008 & 17.542 $\pm$ 0.007 \\
	
\hline
\end{tabular}
\end{minipage}
\end{table*}

For the NTT imaging data, basic reduction (bias subtraction and flat-field correction for the optical; flat-field correction, alignment, illumination and cross-talk correction, sky subtraction and co-addition of images for the NIR data) was done using the purpose-built {\sc python}-based PESSTO reduction pipeline \citep{pessto2} that makes use of Image Reduction and Analysis Facility ({\sc iraf}\footnote{IRAF is distributed by the National Optical Astronomy Observatories, which are operated by the Association of Universities for Research in Astronomy, Inc., under cooperative agreement with the National Science Foundation.}) tasks through {\sc pyraf} and the Source Extractor \citep{sext} and SWarp \citep{swarp} programs. Imaging data from PROMPT1 and TRAPPIST were reduced using similar purpose-built pipelines for those telescopes. The LCOGT imaging data were reduced using a {\sc python}-based custom pipeline by the LCOGT SN team that employs standard {\sc pyraf} and SWarp procedures.

Instrument-specific colour terms for NTT/EFOSC2 were determined using observations of several standard star fields from \citet{std} observed in the photometric nights of 2013 August 29, October 2, November 3 and 2014 January 1. The photometric zero-point calibration of the images was done using several local stars in the field, which were in turn calibrated using the aforementioned standard fields for the optical images, or using the Two Micron All Sky Survey\footnote{http://www.ipac.caltech.edu/2mass/index.html} \citep[2MASS;][]{2mass} for the NIR images. Eight 2MASS stars were present in the field\footnote{2MASS codes 02451042-5543567, 02451550-5545066, 02450981-5545079, 02450060-5545298, 02445765-5545367, 02445641-5544268, 02445894-5544103 and 02445501-5543182.}. Colour terms and zero points for PROMPT1 and TRAPPIST were determined using local stars as well. A \emph{V}-band image of the SN and the field, with the local calibration stars marked and numbered, is shown in Figure 2; the coordinates and the optical magnitudes of the numbered stars are reported in Table 2. The zero points and (small) colour terms of the LCOGT images were determined using 36 local stars in the AAVSO Photometric All-Sky Survey (APASS). The eight 2MASS field stars used for photometric calibration were also used to calibrate the world coordinate system (WCS) of the NTT images using the {\sc iraf} tasks {\sc ccmap} and {\sc ccsetwcs} in the {\sc imcoords} package.

\begin{figure}
\centering
\includegraphics[width=\columnwidth]{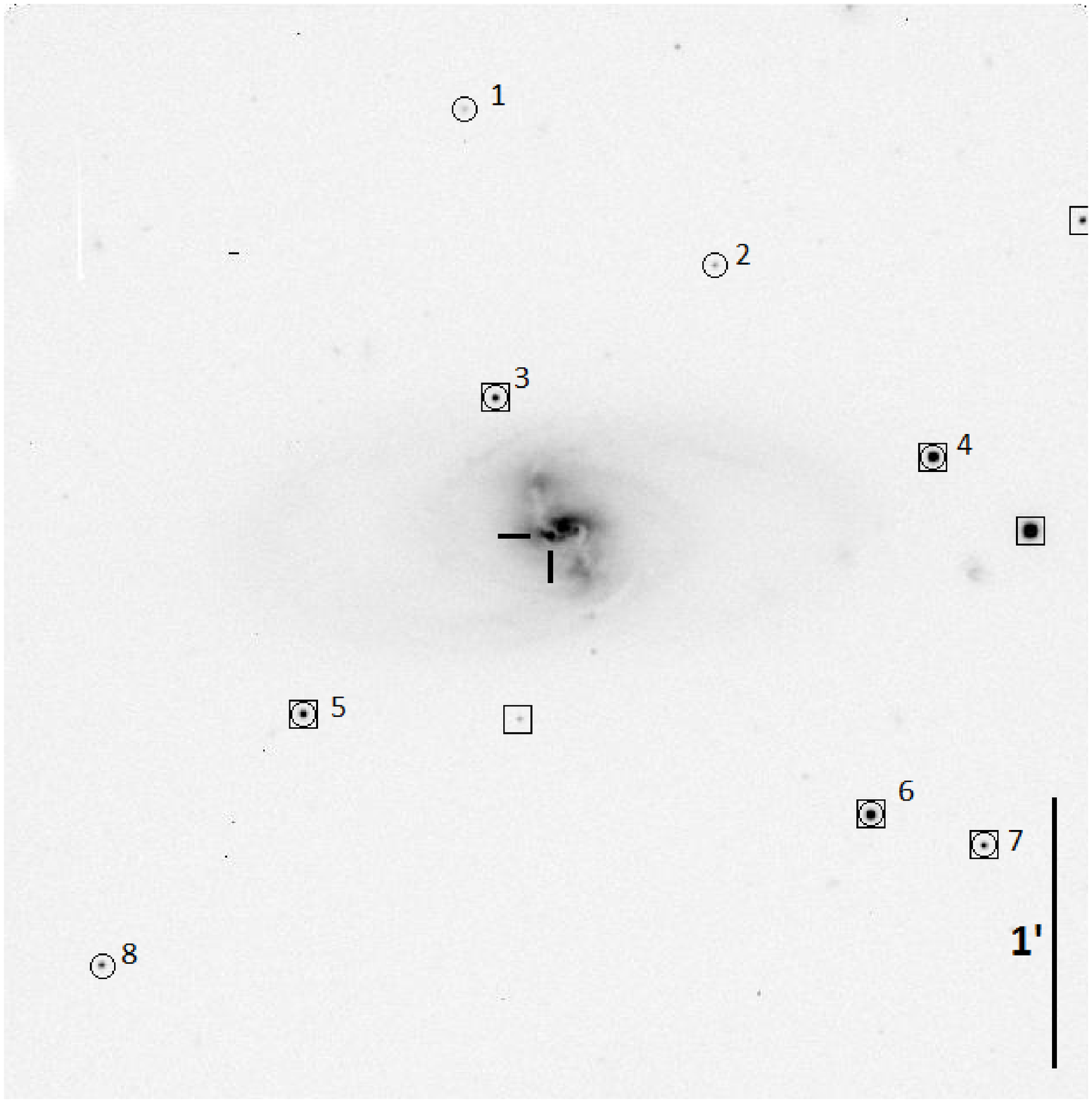}
\caption{SN 2013fc and the surrounding field in the \emph{V} band, as observed with EFOSC2 on 2013 August 29. The local optical calibration stars have been marked with circles, the 2MASS field stars with squares and the SN with tick marks. North is up and east is to the left.}
\end{figure}

In order to obtain more accurate photometric measurements, we performed subtractions of the template images from the earlier imaging data to remove the host galaxy background. The images were first aligned at the sub-pixel level using the {\sc iraf} tasks {\sc geomap} and {\sc geotran}. The subtraction of NTT, PROMPT1 and TRAPPIST images was done with the {\sc isis} 2.2 package using the Optimal Image Subtraction (OIS) method presented by \citet{isis2} and \citet{isis1}. Point spread function (PSF) matching between the template and SN images was performed by deriving a space-varying convolution kernel. Stamps containing a star and some background were identified for the kernel derivation. Roughly ten stars around the galaxy were selected. For each image pair, a few different sets of options -- mainly stamp size and the degree of the polynomial describing spatial variation -- were tested and the best resulting subtracted image (with the most thorough removal of the field stars and the host galaxy) was chosen. Subtractions for the LCOGT data were performed as part of the aforementioned custom LCOGT pipeline, using the {\sc hotpants}\footnote{http://www.astro.washington.edu/users/becker/v2.0/hotpants\\.html} code.

The photometry of the NTT, PROMPT1 and TRAPPIST data was carried out using a PSF fitting procedure based on the {\sc daophot} package in {\sc iraf}. The local stars used for photometric calibration were also used to derive the PSF for each unsubtracted image, which was then fitted to the SN in the subtracted image. Measurement errors were estimated by simulating and fitting nine artificial PSF objects around the position of the SN in the residual image. The reported error is the quadratic sum of this measurement error and the standard error of the mean of the zero-point values derived with the local stars. Most of the {\sc iraf} photometry tasks were run under the {\sc quba} pipeline\footnote{A {\sc python} package specifically designed by Stefano Valenti for SN imaging and spectra reduction. For more details on the pipeline, see \citet{quba}.}. The LCOGT photometry was carried out using the custom LCOGT pipeline, which also performs PSF fitting through {\sc daophot}. 

The NTT spectra were reduced using the PESSTO reduction pipeline. Using {\sc iraf} tasks, they were bias subtracted, flat-fielded, wavelength calibrated with arc lamp exposures, corrected for cosmic rays \citep[using {\sc lacosmic};][]{lacosmic}, and extracted. Sensitivity curves were determined using spectroscopic standard stars observed during the same nights as the spectra and were used for relative flux calibration. The NIR spectra were sky-subtracted, co-added and then calibrated using solar analog telluric standard stars from the Hipparcos catalogue\footnote{http://heasarc.gsfc.nasa.gov/W3Browse/all/hipparcos.html}, observed close to SN 2013fc in both time and airmass. WiFeS data were reduced using the PyWiFeS pipeline \citep{pywifes}, and spectra of the SN were extracted using a custom GUI for selecting `object' and `background' spaxels in the final data cubes. The SALT/RSS spectra were reduced and analysed using the primary reduction PYSALT pipeline \citep{rss2} and the {\sc iraf/twodspec} package in a way described in \citet{rss3}  for another LIRG.

\section{Host galaxy and SN location}

The host galaxy ESO 154-G010, an SBa-type barred spiral galaxy, has a heliocentric redshift of $0.01863 \pm 0.00008$ and a recession velocity of $5586 \pm 24$ km s$^{-1}$ \citep{cat1}. This information was obtained from the NASA/IPAC Extragalactic Database (NED)\footnote{http://ned.ipac.caltech.edu/}. Its IR luminosity has been reported as $\log L_{\textrm{IR}}/L_{\odot} = 10.95$ by \citet{host3}. We adopt a Hubble parameter of $H_{0} = 67.8 \pm 0.9$ km s$^{-1}$ Mpc$^{-1}$, based on the most recent measurements of the cosmic microwave background (CMB) from the \emph{Planck} satellite \citep{planck}. Thus we find the luminosity distance, corrected for Virgo infall, Great Attractor and Shapley supercluster, to be $D_{L} = 83.2 \pm 1.1$ Mpc and the distance modulus $\mu = 34.60 \pm 0.03$ mag. Calibrating with this distance, we in turn obtain $\log L_{\textrm{IR}}/L_{\odot} = 11.05 \pm 0.05$. Therefore the galaxy qualifies as a LIRG ($L_{\textrm{IR}} > 10^{11} L_{\odot}$). As there are no redshift-independent distance measurements for ESO 154-G010 on the NED, we adopt this distance modulus in calculating the absolute magnitudes of SN 2013fc as well. 

Studies of SNe in LIRGs are still somewhat rare \citep[e.g.][]{i17138,ic883,2010op,kangas}. Despite the high star-formation rates (and thus high expected CCSN rates) in these galaxies, few SNe have been discovered in them due to a high extinction in the dusty nuclear/circumnuclear regions where the strongest star formation takes place; roughly 85 per cent of CCSNe have been missed in these regions by optical observations \citep{mattila12}. The star-formation rate of ESO 154-G010, calculated using its far-IR luminosity and Eq. 4 of \citet{sfr}, is 19.42 $M_{\odot}$ yr$^{-1}$, yielding an expected CCSN rate of 0.10 yr$^{-1}$ assuming a Salpeter initial mass function for stars between 0.1 and 125 $M_{\odot}$ and an upper mass limit of 18 $M_{\odot}$ for CCSN progenitors, and 0.14 yr$^{-1}$ for an upper mass limit of 50 $M_{\odot}$.

\begin{table}
\centering
\caption{The FWHM of the narrow H\,$\alpha$ corrected for instrumental broadening, the flux ratios of the narrow [N {\sc ii}]$\lambda6583$/H\,$\alpha$ and [O {\sc iii}]$\lambda5007$/H\,$\beta$ lines, and the Na {\sc i} D line EW, from the WiFeS spectra. The spectra of days 32 and 38 have R = 3000 while the rest have R = 7000.} 
\begin{tabular}[t]{lcccc}
    \hline
        Epoch$^{a}$ & H\,$\alpha$ FWHM & [N {\sc ii}]/H\,$\alpha$ & [O {\sc iii}]/H\,$\beta$ & Na {\sc i} D EW \\
	(days) & (km s$^{-1}$) & & & (\AA~) \\
    \hline
    \hline
	28 & 172 & 0.50 & 0.55 & 7.8 \\
	32 & 187 & 0.51 & 0.57 & 7.7 \\
	38 & 187 & 0.46 & 0.54 & 8.0 \\
	57 & 176 & 0.53 & 0.49 & 7.7 \\
	72 & 183 & 0.52 & 0.57 & 7.2 \\
	93 & 166 & 0.54 & 0.54 & 7.0 \\
	526 & 180 & 0.54 & 0.56 & 7.8 \\
\hline
\end{tabular}
\begin{flushleft}
$^{a}$ Since explosion, in the rest frame of the SN.
\end{flushleft}
\end{table}

We have measured the fluxes and widths of the narrow H\,$\alpha$, H\,$\beta$, [O {\sc iii}] $\lambda\lambda4959,5007$ and [N {\sc ii}] $\lambda\lambda6548,6583$ lines, and the equivalent widths (EW) of the Na {\sc i} D $\lambda\lambda5890,5896$ doublet, from the WiFeS spectra. The 5800 -- 6800 \AA~windows of these spectra (with no host galaxy background subtraction) that include the Na {\sc i} D, H\,$\alpha$, [N {\sc ii}] and [S {\sc ii}] lines are presented in Figure 3, and the measurements are listed in Table 3. The width of the narrow Balmer lines stays fairly constant throughout the observations, with the average full width half maximum (FWHM, corrected for instrumental broadening) of the line being 179 $\pm$ 3 km s$^{-1}$. In addition, we consistently see the [O {\sc iii}], [N {\sc ii}] and [S {\sc ii}] doublets, and the [N {\sc ii}]$\lambda6583$/H\,$\alpha$ (narrow component of H\,$\alpha$) and [O {\sc iii}]$\lambda5007$/H\,$\beta$ flux ratios also stay roughly constant with averages of 0.51 $\pm$ 0.01 and 0.55 $\pm 0.01$ respectively. Finally, the narrow H\,$\alpha$ lines are still visible and unchanged after the SN has faded, in the 526 day spectrum. Therefore we conclude that the narrow emission lines in fact originate from the underlying host galaxy and are not related to the SN itself. The direct consequence of this conclusion is that the classification of the SN as type IIn is questionable, although as mentioned in Section 2, we cannot exclude the presence of early narrow lines due to the other similarities with SN 1998S. Furthermore, any weak narrow lines originating from the SN would be dominated by the galaxy lines.

In addition, we use the narrow-line ratios and Eqs. 2 and 3 \citet{metallicity} to estimate the metallicity at the SN location. Using the [N {\sc ii}]$\lambda6583$/H\,$\alpha$ ratio we have 12 + log(O/H) $\sim 8.7 \pm 0.1$. Using both the [N {\sc ii}]$\lambda6583$/H\,$\alpha$ and [O {\sc iii}]$\lambda5007$/H\,$\beta$ ratios we have the same result. This value is consistent with solar metallicity \citep[12 + log(O/H) = $8.66 \pm 0.05$ according to][]{metallicity2}.

\begin{figure}
\centering
\includegraphics[width=\columnwidth]{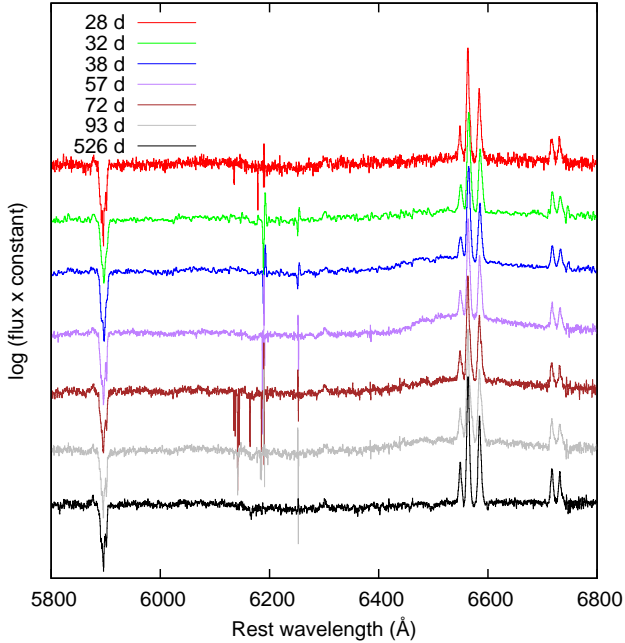}
\caption{The spectra of SN 2013fc from WiFeS, corrected for host galaxy redshift but with no host galaxy background subtraction, zoomed in to show the strong Na {\sc i} D $\lambda\lambda5890,5896$ absorption and the narrow emission lines (H\,$\alpha$, [N {\sc ii}]$\lambda\lambda6548,6583$ and [S {\sc ii}]$\lambda\lambda6717,6730$) originating from the underlying star-forming region. The flux scale is logarithmic for convenience.}
\end{figure}

\begin{figure}
\centering
\includegraphics[width=\columnwidth]{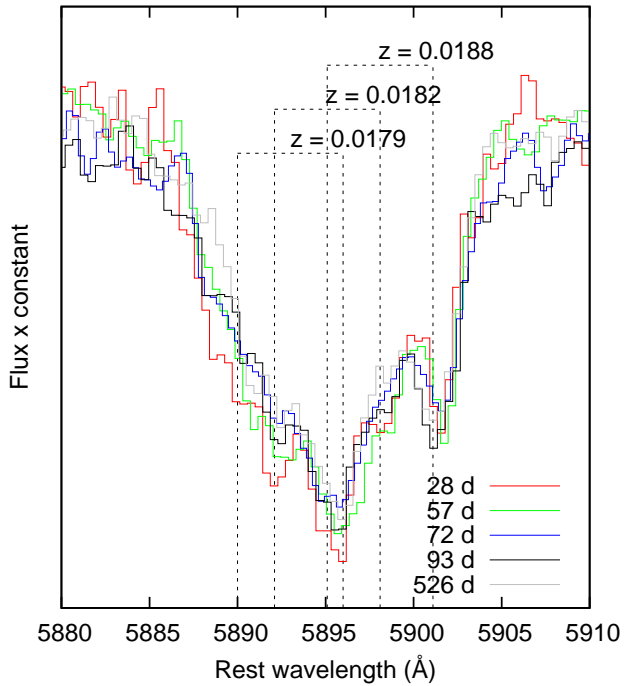}
\caption{The Na {\sc i} D doublet absorption profile in the R=7000 WiFeS spectra of SN 2013fc including the reference spectrum (526 d), illustrating the strong multi-component line. The three doublets at different redshifts have been marked with dashed lines. The redshift correction has been done using $z = 0.0179$. No appreciable evolution can be seen in the line.}
\end{figure}

The location of SN 2013fc is coincident with a circumnuclear ring at a projected distance of $\sim$ 1200 pc from the core. The ring and its clumpy nature can be seen in Figure 1. In order to get accurate positions of the SN and the seemingly coincident bright spot, we have more closely examined the subtraction template images and the first set of subtracted EFOSC2 and SOFI images in the \emph{UBVRIJHK} bands. The RMS error of the coordinate transforms was less than 0$\farcs$1 in all these images; typically from 0$\farcs$05 to 0$\farcs$07. We then took the averages of the resulting coordinates for the SN and the bright spot. This yielded coordinates for SN 2013fc as $\alpha = 02^{\mathrm{h}} 45^{\mathrm{m}} 08\fs988 \pm 0\fs007$, $\delta = -55\degr 44\arcmin 27\farcs37 \pm 0\farcs10$ (equinox J2000.0), which is 2$\farcs$5 east and 1$\farcs$3 south of the host galaxy nucleus as measured in the \emph{K} band. The errors include the RMS uncertainty of the coordinate transform and the standard error of the mean of the position measurements. The bright spot in the template images is at the coordinates $\alpha = 02^{\mathrm{h}} 45^{\mathrm{m}} 08\fs993 \pm 0\fs009$, $\delta = -55\degr 44\arcmin 27\farcs40 \pm 0\farcs11$, which is the average from \emph{RIJHK} band measurements, and which is consistent with the SN coordinates within the uncertainties. The centroid of the bright spot is less clear in the \emph{UBV} band images. The source is visible in the WiFeS template data cube of 2015 January 29, where no sign of the broad H\,$\alpha$ line of the SN can be seen.

Furthermore, as can be seen in Figure 4 and Table 3, the EW and shape of the Na {\sc i} D doublet absorption do not change between the SN spectra and the reference spectrum extracted at this position (where the source of the continuum is the bright coincident source) in the WiFeS data cube of 2015 January 29, after the SN had faded. The profile of the absorption shows a multiple-trough structure, which can be deblended into three doublets of roughly equal strength. These doublets are at the mean redshifts of 0.0179, 0.0182 and 0.0188, measured from the WiFeS spectra; thus the $\lambda5896$ line of the bluest doublet component and the $\lambda5890$ line of the reddest doublet component coincide, creating the aforementioned complex profile. The $z = 0.0179$ doublet also matches the average redshift of the narrow H\,$\alpha$ line from the same spectra, while the redshifts of the two other doublets correspond to velocities of 90 and 270 km s$^{-1}$, respectively, relative to the H\,$\alpha$. This, and the fact that the shape stays the same after the SN fades, provide additional circumstantial evidence that the SN exploded within the bright spot; thus we have also used the redshift 0.0179 for the rest frame correction of the SN spectra (see Section 5). 

As no spectral features of SN 2013fc can be seen in the day 526 reference spectrum, it was used to subtract the background contamination from the SN spectra (see Section 5). We also fitted a stellar population model to this spectrum using the Starlight (SL) spectral synthesis code \citep{starlight}. SL finds the best-fitting overall spectral model as a superposition of individual input single stellar populations (SSP) and outputs the weights required from each SSP separately for light and mass. As input models we used \citet{slmodels} SSPs with Padova 1994 tracks \citep{padova94} and a Salpeter initial mass function. The spaxel size of the WiFeS data cubes is 1$\arcsec$ and the seeing in the night of 2015 January 29, was 1$\farcs$5, which makes it non-trivial to extract a reference spectrum of the coincident source alone; thus we have applied the SL code to spectra extracted with slightly different apertures including the SN location. The resulting best fits show a mixture of different populations with comparable contributions to the total light: a young population $10_{-2}^{+3}$ Myr of age; an intermediate-age population between a few hundred Myr and 1 Gyr; and an old population of $> 1$ Gyr. The metallicity in the best fits was approximately solar or slightly less, consistently with our estimate based on the line ratios. The error for the age of the youngest population has been estimated based on this metallicity range and the age intervals of the Padova isochrones. The age of the young population, although by no means an accurate tool due to the spatial resolution, can be used to very roughly assess the progenitor mass (see Section 6).

One of the background extractions (the differences between them are small) and the best SL fit to it are shown in the upper panel of Figure 5, along with the nuclear spectrum of the galaxy extracted from the same data cube. The lower panel contains the wavelength region around the H\,$\alpha$ line from the higher S/N 378 d SALT spectrum, extracted at the nucleus, the northern lobe of the galaxy and the part of the circumnuclear ring located between them. The SALT spectra of the nucleus and ring areas (the latter very similar to the SN location spectrum) were also fitted with SL. The results show essentially an old evolved galaxy, with additional signs of star formation around 500 Myr to 1 Gyr ago potentially linked to some past interaction or minor merger event, and then a very young 5-10 Myr age current star formation episode. The spectra show that the ring around the nucleus is the brightest emission line component.

The FWHM of the H\,$\alpha$ line in the nuclear spectrum is $\sim$ 260 km s$^{-1}$ (consistent with a Seyfert 2 galaxy), and the line flux ratios, [N {\sc ii}]$\lambda$6583/H$\alpha \sim 0.8$ and [O {\sc iii}]$\lambda$5007/H\,$\beta \sim 0.9$, are consistent with a composite of an active galactic nucleus (AGN) and a star-forming galaxy according to the classification scheme of \citet{host2}. However, the [S {\sc ii}]$\lambda\lambda$6717,6730/H\,$\alpha$ flux ratio ($\sim 0.4$) is more consistent with a star-forming region. As mentioned, the FWHM of the Balmer lines at the location of the SN is roughly 180 km s$^{-1}$; this is inconsistent with the line width at a typical H {\sc ii} region in a spiral galaxy (normally on the order of some tens of km s$^{-1}$) but consistent with the extended narrow-line region (ENLR) of a Seyfert galaxy nucleus, extending to the circumnuclear ring \citep[e.g.][]{seyfert2}. However, the width is also consistent with the turbulence in a LIRG, caused by poweful star-formation and/or the kinematic effects of galaxy interaction \citep{lirg1}. Furthermore, line-fitting to the H\,$\alpha$ profile at the nuclear spectrum also shows a broad component with an asymmetric shape more consistent with gas outflows than the broad-line region of an AGN. The [N {\sc ii}]$\lambda$6583/H\,$\alpha$ flux ratios in the circumnuclear ring and the northern lobe are also consistent with shock ionization of gas due to outflows \citep{lirg2}, further supported by the low nuclear [S {\sc ii}]$\lambda\lambda$6717,6730/H\,$\alpha$ flux ratio that is inconsistent with an AGN. Thus we conclude that the presence of an AGN is not likely.

\begin{figure}
\centering
\includegraphics[width=\columnwidth]{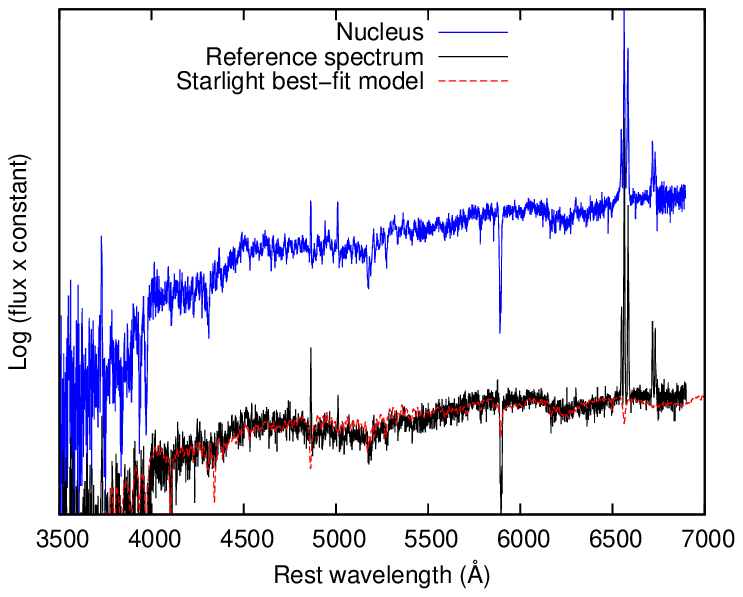}
\includegraphics[width=\columnwidth]{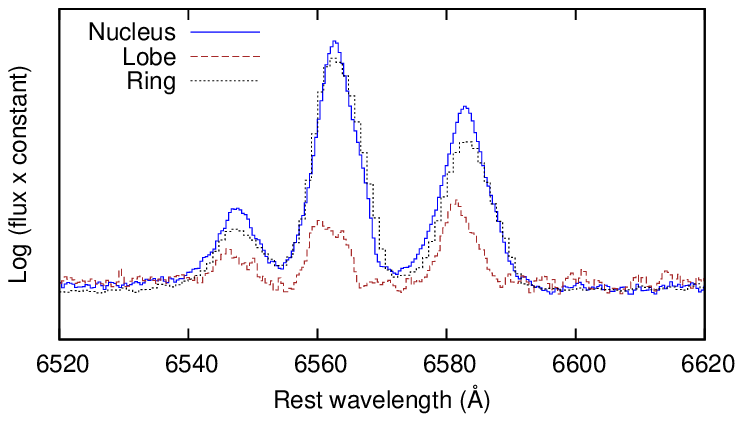}
\caption{Upper panel: The nuclear spectrum (blue), and a reference spectrum (black) at the SN location extracted from the 526 d data cube, with the best Starlight stellar population fit (red). The scaling between the nuclear and reference spectra is arbitrary. Lower panel: The H\,$\alpha$ and [N {\sc ii}] profiles of the R=3000 SALT spectra of the nucleus (blue solid line) and the northern lobe (brown dashed line) and the R=1000 spectrum of the circumnuclear ring (black dotted line).}
\end{figure}

\subsection{Extinction}

Having determined that the narrow hydrogen Balmer lines in the spectrum of SN 2013fc in fact originate from the host galaxy, we also estimated the extinction along the line of sight to SN 2013fc using the Balmer decrement. Subtracting the SL stellar population model fit from the WiFeS reference spectrum, we remove the stellar absorption features and the continuum, and are left with a pure emission spectrum of the underlying H {\sc ii} region. Using the extinction law of \citet{ext3}, we obtain an estimate of $A_{V} = 2.8 \pm 0.2$ mag as an average from the different extractions. This value has been obtained assuming that the intrinsic Balmer decrement follows the standard case B recombination.

The EW of the Na {\sc i} D absorption is one of the strongest observed for SNe, with an average value of 7.6 $\pm$ 0.2 \AA. As can be seen in Figure 5, the absorption in the nucleus is also strong, with an EW of 7.1 \AA. The three-component structure of the absorption indicates the presence of at least three strongly absorbing clouds along the line of sight. The EW of the Na {\sc i} D doublet has been used to estimate the extinction to many SNe; however, different relations for the extinction vs. the EW exist, and at values as high as this there is a large scatter in the resulting $A_{V}$ values \citep{naid}. \citet{lirg3} also showed that, while large Na {\sc i} D EWs are commonly observed in LIRGs, reaching values of up to 10 \AA~or even more, there is a large scatter in the corresponding $A_{V}$ values. Thus this method has not been used here.

The possibility exists that an extinction estimated using the Balmer decrement would be erroneous; the SN could actually be behind or in front of the line-emitting region we use in the calculation. To independently estimate the host galaxy extinction of SN 2013fc we used a method of multi-band template light curve comparison with a simultaneous $\chi^{2}$ fit, by comparing the \emph{BVRI} light curves of SN 2013fc to those of SN 1998S (see Section 4 for the photometry and a comparison to SN 1998S). It was assumed that the evolution (but not necessarily the absolute magnitudes) of the \emph{BVRI} light curves, and the optical colours of the events, would be intrinsically similar. For details of the method, see \citet{2010op, 2005at}. A total line-of-sight extinction of $A_{V} = 0.68$ mag was adopted for SN 1998S \citep{1998S2}. The shapes of the \emph{JHK} light curves of the events are somewhat different, possibly due to different contributions from dust emission (and/or a late-time NIR emission still being present in our template images; see Section 4), and they were disregarded so as to not introduce a bias to the fitting process. We obtain the result $A_{V} = 3.0$ mag with the lowest $\chi^{2}$ value assuming the standard Cardelli extinction law. Furthermore, a $\chi^{2}$ test was performed between spectra of SN 2013fc (where host galaxy contamination was subtracted away) and SN 1998S at early epochs where, by coincidence, spectra for both SNe exist, $\sim$2 and $\sim$12 days after maximum. From these fits, with the assumption that the spectral shapes were intrinsically similar (see Section 5), we obtain the result $A_{V} = 3.0$ mag as well. As noted above, the Na {\sc i} D absorption line profile is also the same whether measured against the SN or the coincident source, suggesting that the SN is indeed associated with the emission region. With reasonable agreement from these estimates, we adopt the average value of the three, $A_{V} = 2.9 \pm 0.2$ mag, for the total line-of-sight extinction to SN 2013fc. The error estimate here is simply the error from the Balmer-decrement calculation. The Galactic extinction, determined using the dust maps of \citet{ext5}, is $A_{V,Gal} = 0.082$ mag. This value is included in the total extinction value.

This extinction estimation has made use of the Cardelli et al. (1989) extinction law with $R_{V} = 3.1$. However, it is possible that the extinction law in the circumnuclear region of a LIRG such as ESO 154-G010 could be different. We have also performed the $\chi^{2}$ fitting between SN 2013fc and SN 1998S using different values of $R_{V}$; the parameters and results of these fits are presented in Table 4. Furthermore, \citet{ext6} have derived an attenuation law for actively star-forming galaxies with dust content, which has been used in some cases as an extinction law for e.g. SNe in starburst galaxies \citep[e.g.][]{2010op}. Using this law, we would obtain an extinction of $A_{V} = 3.6$ mag, and the NIR colours of SN 2013fc would be more similar to those of SN 1998S. However, this law is not a clear-cut extinction law and not necessarily applicable to point sources such as SNe; and due to the somewhat different shapes of the NIR light curves and the difference in luminosity vs. SN 1998S, expecting the NIR (dust) contribution to be the same is not well justified. As there does not seem to be a consensus in the literature as to what value of $R_{V}$ to use in this case, we have conservatively settled on the commonly used Cardelli extinction law with $R_{V} = 3.1$.

\begin{table}
\centering
\caption{The parameters ($R_{V}$ and extinction law) and results (extinction $A_{V}$, the epoch of discovery $t_{0}$ assuming identical rise times, brightness difference $\Delta M$) of the SN 2013fc vs. SN 1998S \emph{BVRI} light curve $\chi^{2}$ fits.} 
\begin{tabular}[t]{lccccc}
    \hline
        $R_{V}$ & Extinction & $A_{V}$ & $t_{0}$ & $\Delta M$ & Reduced $\chi^{2}$ \\
	 & law & (mag) & (d) & (mag) & \\
    \hline
    \hline
	2.1 & Cardelli & 2.2 & 8 & 0.2 & 10.2 \\
	2.6 & Cardelli & 2.7 & 7 & 0.6 & 10.3 \\
	3.1 & Cardelli & 3.0 & 7 & 0.9 & 11.1 \\
	3.6 & Cardelli & 3.4 & 8 & 1.3 & 12.3 \\
	4.1 & Cardelli & 3.6 & 8 & 1.5 & 13.9 \\
	4.6 & Cardelli & 4.0 & 8 & 1.8 & 15.6 \\
	5.1 & Cardelli & 4.2 & 7 & 2.0 & 16.8 \\
    \hline
	4.05 & Calzetti & 3.6 & 7 & 1.4 & 13.4 \\
\hline
\end{tabular}
\end{table}

\section{Photometry}

\begin{table*}
\begin{minipage}{160mm}
\centering
\caption{The apparent \emph{UBVRI} magnitudes of SN 2013fc. The \emph{R} and \emph{I} magnitudes are reported in the Johnson-Cousins system. Unfiltered data are marked with an asterisk (*). The epoch is in days since explosion, in the rest frame of the SN.}
\begin{tabular}[t]{lccccccc}
    \hline
        JD & Epoch & \emph{U} & \emph{B} & \emph{V} & \emph{R} & \emph{I} & Telescope \\
	(2400000+) & (days) & (mag) & (mag) & (mag) & (mag) & (mag) & \\
    \hline
    \hline
	56524.7 & 7.9 & - & - & - & 16.71 $\pm$ 0.09 & - & PROMPT1*\\
	56525.7 & 8.8 & - & - & - & 16.86 $\pm$ 0.11 & - & PROMPT1*\\
	56526.7 & 9.8 & - & - & - & 16.70 $\pm$ 0.06 & - & PROMPT1*\\
	56530.8 & 13.9 & - & 18.02 $\pm$ 0.04 & 17.20 $\pm$ 0.03 & - & - & LCOGT 1m0-4\\
	56531.8 & 14.8 & - & 18.04 $\pm$ 0.05 & 17.24 $\pm$ 0.03 & - & - & LCOGT 1m0-4\\
	56531.9 & 14.9 & - & - & - & 16.67 $\pm$ 0.09 & - & PROMPT1\\	
	56532.9 & 15.9 & - & 18.14 $\pm$ 0.04 & - & - & - & LCOGT 1m0-4\\
	56533.9 & 16.9 & - & 18.15 $\pm$ 0.04 & 17.27 $\pm$ 0.06 & - & - & LCOGT 1m0-5\\
	56534.9 & 17.9 & 18.58 $\pm$ 0.02 & 18.32 $\pm$ 0.02 & 17.32 $\pm$ 0.01 & 16.73 $\pm$ 0.03 & 16.21 $\pm$ 0.01 & NTT\\
	56535.9 & 18.8 & - & 18.47 $\pm$ 0.16 & - & - & - & LCOGT 1m0-9\\
	56535.9 & 18.9 & - & 18.35 $\pm$ 0.47 & 16.98 $\pm$ 0.16 & 16.88 $\pm$ 0.17 & - & PROMPT1\\		
	56536.9 & 19.8 & - & 18.29 $\pm$ 0.11 & 17.46 $\pm$ 0.10 & 16.63 $\pm$ 0.09 & 16.30 $\pm$ 0.14 & PROMPT1\\
	56537.8 & 20.7 & - & 18.42 $\pm$ 0.23 & 17.31 $\pm$ 0.06 & 16.74 $\pm$ 0.04 & 16.24 $\pm$ 0.05 & PROMPT1\\
	56538.6 & 21.5 & - & 18.46 $\pm$ 0.06 & 17.44 $\pm$ 0.05 & - & - & LCOGT 1m0-12\\
	56538.9 & 21.8 & - & 18.30 $\pm$ 0.18 & 17.43 $\pm$ 0.14 & 16.77 $\pm$ 0.05 & - & PROMPT1\\		
	56539.8 & 22.7 & - & - & - & 16.89 $\pm$ 0.08 & 16.39 $\pm$ 0.06 & PROMPT1\\
	56540.6 & 23.4 & - & 18.56 $\pm$ 0.08 & 17.77 $\pm$ 0.07 & - & - & LCOGT 1m0-12\\
	56542.6 & 25.4 & - & 18.81 $\pm$ 0.07 & - & - & - & LCOGT 1m0-13\\
	56543.9 & 26.7 & 19.53 $\pm$ 0.10 & 18.90 $\pm$ 0.03 & 17.71 $\pm$ 0.02 & 17.04 $\pm$ 0.01 & 16.41 $\pm$ 0.03 & NTT\\
	56545.1 & 27.9 & - & 18.75 $\pm$ 0.11 & 17.65 $\pm$ 0.06 & - & - & LCOGT 1m0-3\\
	56545.8 & 28.6 & - & 18.72 $\pm$ 0.17 & 17.75 $\pm$ 0.11 & 17.01 $\pm$ 0.04 & 16.47 $\pm$ 0.08 & PROMPT1\\
	56546.8 & 29.6 & - & - & 17.65 $\pm$ 0.06 & - & - & PROMPT1\\
	56547.2 & 30.0 & - & 18.84 $\pm$ 0.09 & 17.84 $\pm$ 0.05 & - & - & LCOGT 1m0-11\\
	56547.9 & 30.7 & - & - & - & 17.10 $\pm$ 0.15 & - & PROMPT1\\
	56548.9 & 31.6 & - & - & - & 17.36 $\pm$ 0.11 & 16.38 $\pm$ 0.11 & PROMPT1\\
	56552.8 & 35.5 & - & 19.48 $\pm$ 0.16 & 18.07 $\pm$ 0.07 & - & - & LCOGT 1m0-4\\
	56554.5 & 37.2 & - & 19.21 $\pm$ 0.11 & 18.15 $\pm$ 0.07 & - & - & LCOGT 1m0-12,13\\
	56554.9 & 37.5 & - & 19.19 $\pm$ 0.22 & 18.06 $\pm$ 0.20 & - & - & PROMPT1\\
	56556.8 & 39.4 & - & 19.36 $\pm$ 0.14 & 18.22 $\pm$ 0.06 & - & - & LCOGT 1m0-5\\
	56559.6 & 42.1 & - & 19.76 $\pm$ 0.14 & 18.53 $\pm$ 0.09 & - & - & LCOGT 1m0-13\\
	56559.8 & 42.3 & - & 19.72 $\pm$ 0.09 & 18.36 $\pm$ 0.05 & - & - & LCOGT 1m0-4\\
	56559.9 & 42.4 & - & - & 17.97 $\pm$ 0.09 & 17.40 $\pm$ 0.07 & - & PROMPT1\\
	56560.9 & 43.4 & - & - & - & 17.49 $\pm$ 0.07 & 16.89 $\pm$ 0.06 & PROMPT1\\
	56561.2 & 43.7 & - & 19.59 $\pm$ 0.14 & 18.30 $\pm$ 0.05 & - & - & LCOGT 1m0-3\\
	56563.2 & 45.7 & - & 19.74 $\pm$ 0.11 & 18.30 $\pm$ 0.05 & - & - & LCOGT 1m0-3\\
	56566.8 & 49.2 & - & 20.02 $\pm$ 0.08 & 18.45 $\pm$ 0.06 & 17.38 $\pm$ 0.03 & 16.84 $\pm$ 0.03 & TRAPPIST\\
	56568.8 & 51.2 & 21.24 $\pm$ 0.21 & 19.88 $\pm$ 0.05 & 18.42 $\pm$ 0.03 & 17.62 $\pm$ 0.02 & 16.98 $\pm$ 0.05 & NTT\\
	56572.5 & 54.8 & - & 19.74 $\pm$ 0.27 & 18.73 $\pm$ 0.14 & - & - & LCOGT 1m0-13\\
	56574.5 & 56.8 & - & 19.93 $\pm$ 0.12 & 18.65 $\pm$ 0.07 & - & - & LCOGT 1m0-12\\
	56576.8 & 59.0 & - & 20.50 $\pm$ 0.11 & 18.66 $\pm$ 0.05 & 17.57 $\pm$ 0.03 & 17.01 $\pm$ 0.03 & TRAPPIST\\
	56579.5 & 61.7 & - & 20.37 $\pm$ 0.26 & 19.02 $\pm$ 0.08 & - & - & LCOGT 1m0-12\\
	56581.8 & 64.0 & - & 20.53 $\pm$ 0.10 & 18.68 $\pm$ 0.02 & 17.84 $\pm$ 0.02 & 17.22 $\pm$ 0.02 & NTT\\
	56582.7 & 64.9 & - & - & 18.87 $\pm$ 0.11 & - & - & LCOGT 1m0-5\\
	56585.7 & 67.8 & - & 20.32 $\pm$ 0.14 & 19.09 $\pm$ 0.07 & - & - & LCOGT 1m0-4,5\\
	56589.4 & 71.4 & - & - & 19.64 $\pm$ 0.31 & - & - & LCOGT 1m0-13\\
	56589.8 & 71.8 & - & 21.08 $\pm$ 0.13 & 19.29 $\pm$ 0.05 & 18.21 $\pm$ 0.04 & 17.54 $\pm$ 0.06 & NTT\\
	56593.7 & 75.6 & - & - & 19.43 $\pm$ 0.20 & - & - & LCOGT 1m0-4\\
	56594.7 & 76.6 & - & - & 19.70 $\pm$ 0.17 & - & - & LCOGT 1m0-5\\
	56599.7 & 81.6 & - & 21.03 $\pm$ 0.41 & 20.56 $\pm$ 0.31 & - & - & LCOGT 1m0-9\\
	56600.7 & 82.5 & - & 21.45 $\pm$ 0.11 & 20.15 $\pm$ 0.06 & 19.02 $\pm$ 0.07 & 18.40 $\pm$ 0.08 & NTT\\
	56600.7 & 82.5 & - & 20.89 $\pm$ 0.23 & 20.23 $\pm$ 0.16 & - & - & LCOGT 1m0-9\\
	56601.5 & 83.3 & - & 20.99 $\pm$ 0.46 & 20.46 $\pm$ 0.21 & - & - & LCOGT 1m0-12\\
	56607.6 & 89.3 & - & - & 20.70 $\pm$ 0.25 & 19.02 $\pm$ 0.09 & 18.60 $\pm$ 0.17 & TRAPPIST\\
	56609.6 & 91.3 & - & - & 20.27 $\pm$ 0.32 & - & - & LCOGT 1m0-4\\
	56609.9 & 91.6 & - & - & 20.51 $\pm$ 0.13 & 19.28 $\pm$ 0.05 & 18.62 $\pm$ 0.05 & NTT\\
	56615.6 & 97.2 & - & - & 20.14 $\pm$ 0.25 & - & - & LCOGT 1m0-5\\
	56617.3 & 98.8 & - & - & 20.65 $\pm$ 0.30 & - & - & LCOGT 1m0-12\\
	56619.7 & 101.2 & - & 22.14 $\pm$ 0.20 & 20.78 $\pm$ 0.11 & 19.45 $\pm$ 0.05 & - & NTT\\
	56623.6 & 105.0 & - & - & 20.91 $\pm$ 0.56 & - & - & LCOGT 1m0-4\\
	56630.3 & 111.6 & - & - & 20.31 $\pm$ 0.23 & - & - & LCOGT 1m0-12\\
	56635.5 & 116.7 & - & - & 20.93 $\pm$ 0.24 & 19.50 $\pm$ 0.07 & 19.08 $\pm$ 0.19 & NTT\\
	56640.7 & 121.8 & - & - & 20.73 $\pm$ 0.33 & - & - & LCOGT 1m0-9\\
	56682.6 & 163.0 & - & 23.28 $\pm$ 0.43 & 21.68 $\pm$ 0.15 & 20.53 $\pm$ 0.07 & 20.04 $\pm$ 0.09 & NTT\\
\hline
\end{tabular}
\end{minipage}
\end{table*}
\begin{table*}
\begin{minipage}{115mm}
\centering
\caption{The apparent \emph{gri} magnitudes of SN 2013fc.}
\begin{tabular}[t]{lcccccc}
    \hline
       	JD & Epoch\footnote{Since explosion, in the rest frame of the SN.} & \emph{g} & \emph{r} & \emph{i} & Telescope\\
	(2400000+) & (days) & (mag) & (mag) & (mag) & \\
    \hline
    \hline
	56530.8 & 13.9 & 17.67 $\pm$ 0.03 & 16.90 $\pm$ 0.02 & 16.71 $\pm$ 0.03 & LCOGT 1m0-4 \\
	56531.8 & 14.8 & 17.68 $\pm$ 0.03 & 16.88 $\pm$ 0.03 & 16.66 $\pm$ 0.03 & LCOGT 1m0-4 \\
	56532.9 & 15.9 & - & - & 16.68 $\pm$ 0.04 & LCOGT 1m0-4 \\
	56533.9 & 16.9 & 17.77 $\pm$ 0.04 & 16.89 $\pm$ 0.05 & 16.64 $\pm$ 0.03 & LCOGT 1m0-5 \\
	56535.9 & 18.8 & 17.77 $\pm$ 0.04 & 17.28 $\pm$ 0.06 & 16.64 $\pm$ 0.07 & LCOGT 1m0-9 \\
	56538.6 & 21.5 & 18.13 $\pm$ 0.05 & 16.95 $\pm$ 0.04 & 16.83 $\pm$ 0.06 & LCOGT 1m0-12 \\
	56540.6 & 23.4 & 18.20 $\pm$ 0.06 & 17.07 $\pm$ 0.04 & 16.75 $\pm$ 0.05 & LCOGT 1m0-12 \\
	56542.6 & 25.4 & 18.36 $\pm$ 0.08 & 17.15 $\pm$ 0.05 & 16.83 $\pm$ 0.05 & LCOGT 1m0-13 \\
	56545.1 & 27.9 & 18.50 $\pm$ 0.06 & 17.28 $\pm$ 0.05 & 16.92 $\pm$ 0.04 & LCOGT 1m0-3 \\
	56547.2 & 30.0 & 18.67 $\pm$ 0.07 & 17.41 $\pm$ 0.04 & 17.04 $\pm$ 0.05 & LCOGT 1m0-11 \\
	56552.8 & 35.5 & - & 17.37 $\pm$ 0.05 & 17.34 $\pm$ 0.07 & LCOGT 1m0-4 \\
	56554.5 & 37.2 & 19.01 $\pm$ 0.08 & 17.58 $\pm$ 0.03 & 17.26 $\pm$ 0.04 & LCOGT 1m0-12,13 \\
	56556.8 & 39.4 & 18.93 $\pm$ 0.08 & 17.63 $\pm$ 0.05 & 17.22 $\pm$ 0.06 & LCOGT 1m0-5 \\
	56559.6 & 42.1 & 19.15 $\pm$ 0.06 & 17.61 $\pm$ 0.04 & - & LCOGT 1m0-13 \\
	56559.8 & 42.3 & 19.13 $\pm$ 0.04 & 17.62 $\pm$ 0.03 & 17.39 $\pm$ 0.05 & LCOGT 1m0-4 \\
	56561.2 & 43.7 & 19.29 $\pm$ 0.05 & 17.73 $\pm$ 0.04 & 17.52 $\pm$ 0.06 & LCOGT 1m0-3 \\
	56563.2 & 45.7 & - & - & 17.35 $\pm$ 0.05 & LCOGT 1m0-3 \\
	56572.5 & 54.8 & 19.56 $\pm$ 0.12 & 17.73 $\pm$ 0.06 & 17.44 $\pm$ 0.08 & LCOGT 1m0-13 \\
	56574.5 & 56.8 & 19.61 $\pm$ 0.07 & 17.93 $\pm$ 0.04 & 17.58 $\pm$ 0.05 & LCOGT 1m0-12 \\
	56579.5 & 61.7 & 19.96 $\pm$ 0.09 & 17.82 $\pm$ 0.04 & 17.77 $\pm$ 0.06 & LCOGT 1m0-12 \\
	56582.7 & 64.9 & 20.22 $\pm$ 0.10 & 17.99 $\pm$ 0.04 & 17.75 $\pm$ 0.06 & LCOGT 1m0-5 \\
	56585.7 & 67.8 & 19.99 $\pm$ 0.08 & 18.10 $\pm$ 0.03 & 17.92 $\pm$ 0.05 & LCOGT 1m0-4,5 \\
	56589.4 & 71.4 & - & 18.17 $\pm$ 0.06 & 18.08 $\pm$ 0.07 & LCOGT 1m0-13 \\
	56593.7 & 75.6 & - & 18.62 $\pm$ 0.04 & 18.28 $\pm$ 0.08 & LCOGT 1m0-4 \\
	56594.7 & 76.6 & - & 18.62 $\pm$ 0.04 & 18.43 $\pm$ 0.09 & LCOGT 1m0-5 \\
	56599.7 & 81.6 & 21.31 $\pm$ 0.30 & 19.18 $\pm$ 0.18 & 19.13 $\pm$ 0.16 & LCOGT 1m0-9 \\
	56600.7 & 82.5 & 21.83 $\pm$ 0.35 & 19.06 $\pm$ 0.04 & 19.08 $\pm$ 0.15 & LCOGT 1m0-9 \\
	56601.5 & 83.3 & - & 19.24 $\pm$ 0.10 & 19.08 $\pm$ 0.18 & LCOGT 1m0-12 \\
	56604.7 & 86.4 & - & 19.20 $\pm$ 0.08 & 19.80 $\pm$ 0.19 & LCOGT 1m0-9 \\
	56609.7 & 91.3 & 21.61 $\pm$ 0.20 & 19.40 $\pm$ 0.11 & 19.54 $\pm$ 0.14 & LCOGT 1m0-4 \\
	56613.6 & 95.2 & - & 19.28 $\pm$ 0.11 & 19.16 $\pm$ 0.21 & LCOGT 1m0-9 \\
	56615.6 & 97.2 & - & 19.25 $\pm$ 0.08 & 19.21 $\pm$ 0.14 & LCOGT 1m0-5 \\
	56617.4 & 98.9 & - & 19.23 $\pm$ 0.10 & 19.52 $\pm$ 0.22 & LCOGT 1m0-12 \\
	56623.6 & 105.0 & - & 19.32 $\pm$ 0.07 & 19.40 $\pm$ 0.16 & LCOGT 1m0-4 \\
	56630.3 & 111.6 & - & 19.19 $\pm$ 0.09 & 19.47 $\pm$ 0.16 & LCOGT 1m0-12 \\
	56640.7 & 121.8 & - & 19.11 $\pm$ 0.14 & 19.99 $\pm$ 0.16 & LCOGT 1m0-9 \\
\hline
\end{tabular}
\end{minipage}
\end{table*}
\begin{table*}
\begin{minipage}{115mm}
\centering
\caption{The apparent NIR magnitudes of SN 2013fc. All the data are from NTT/SOFI.}
\begin{tabular}[t]{lccccc}
    \hline
       	JD & Epoch\footnote{Since explosion, in the rest frame of the SN.} & \emph{J} & \emph{H} & \emph{K}\\
	(2400000+) & (days) & (mag) & (mag) & (mag) \\
    \hline
    \hline
	56569.7 & 52.1 & 15.91 $\pm$ 0.03 & 15.33 $\pm$ 0.03 & 14.80 $\pm$ 0.04\\
	56579.9 & 62.1 & 16.05 $\pm$ 0.04 & 15.42 $\pm$ 0.03 & 14.91 $\pm$ 0.04\\
	56591.8 & 73.8 & 16.21 $\pm$ 0.04 & 15.59 $\pm$ 0.05 & 15.12 $\pm$ 0.07\\
	56599.6 & 81.4 & 16.73 $\pm$ 0.04 & 15.90 $\pm$ 0.05 & 15.29 $\pm$ 0.04\\
	56609.9 & 91.6 & 16.94 $\pm$ 0.03 & 15.91 $\pm$ 0.04 & 15.28 $\pm$ 0.06\\
	56636.6 & 117.8 & 17.42 $\pm$ 0.07 & 16.49 $\pm$ 0.08 & 15.94 $\pm$ 0.07\\
	56660.5 & 141.3 & 18.10 $\pm$ 0.10 & 17.69 $\pm$ 0.11 & 17.05 $\pm$ 0.10\\
	56682.6 & 163.0 & 18.22 $\pm$ 0.10 & 17.53 $\pm$ 0.16 & 16.97 $\pm$ 0.07\\
	56724.5 & 204.2 & 18.91 $\pm$ 0.21 & 18.03 $\pm$ 0.17 & 17.30 $\pm$ 0.22\\
\hline
\end{tabular}
\end{minipage}
\end{table*}

\begin{figure*}
\centering
\begin{minipage}{160mm}
\includegraphics[width=\columnwidth]{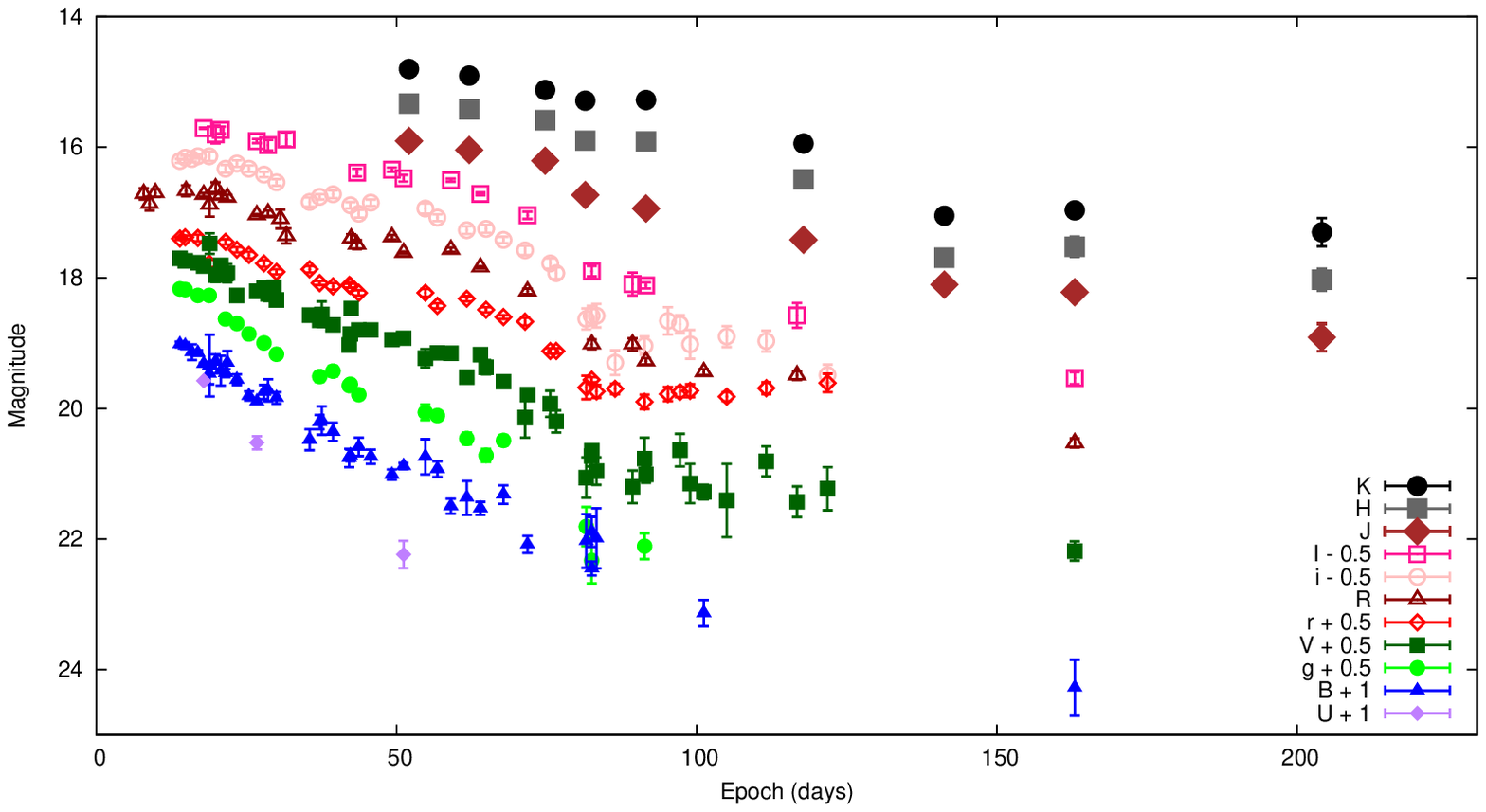}
\includegraphics[width=\columnwidth]{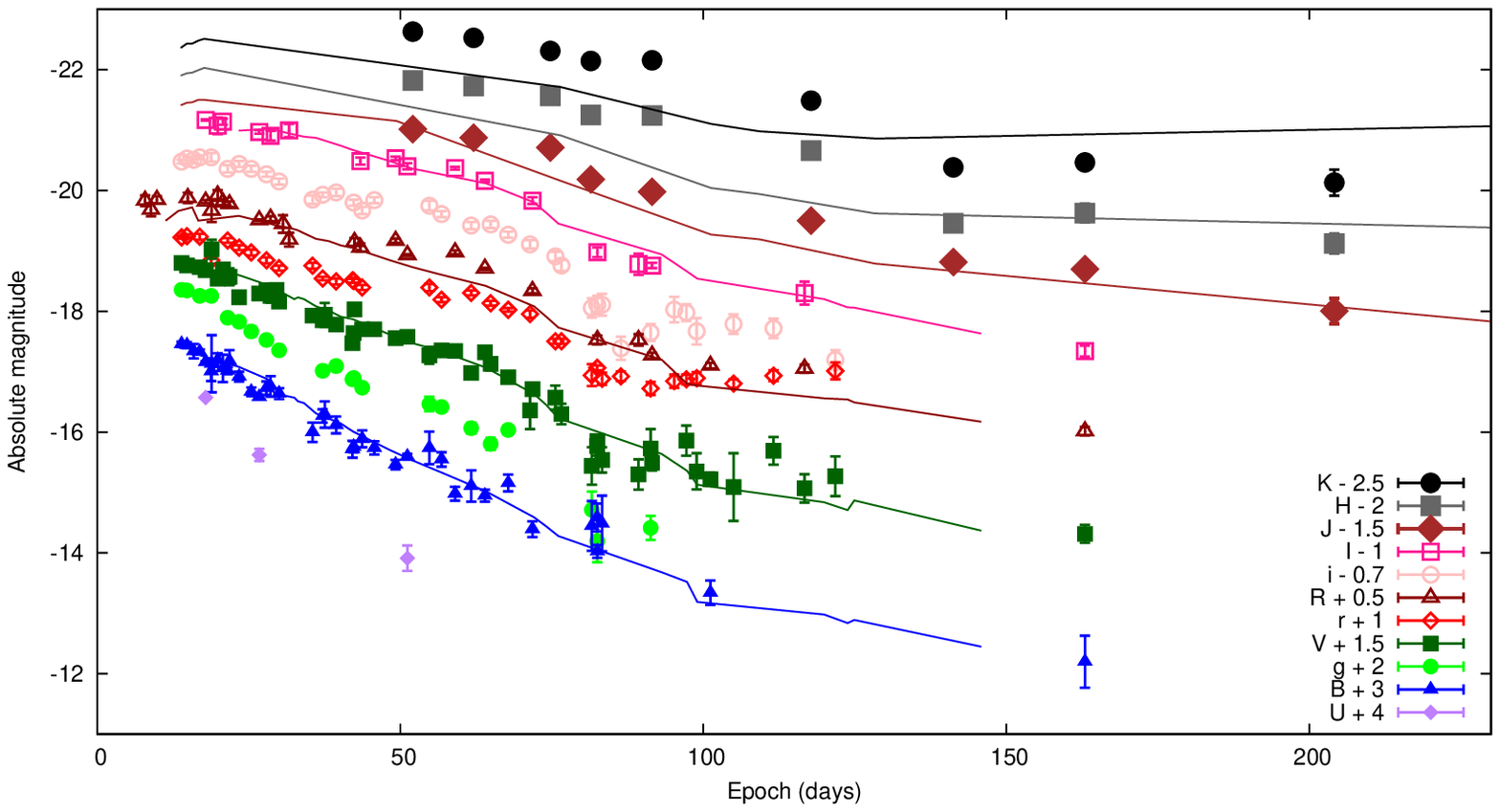}
\caption{Upper panel: The apparent optical and NIR light curves of SN 2013fc. Lower panel: The absolute optical and NIR light curves of SN 2013fc (dots), and the \emph{BVRIJHK} light curves of SN 1998S \citep[lines, with a -0.9 mag offset;][]{1998S2} to illustrate the optical similarity between the two events.}
\end{minipage}
\end{figure*}

\subsection{The light curves}

The apparent \emph{UBVRI}, \emph{gri} and \emph{JHK} magnitudes of SN 2013fc are listed in Tables 5, 6 and 7, respectively. The apparent light curves in all bands are presented in the upper panel of Figure 6. In the lower panel, we show the absolute light curves of SN 2013fc and those of SN 1998S, with an offset of -0.9 mag. Early unfiltered data from PROMPT1 have been approximated as \emph{R}-band points by using the \emph{R} magnitudes of the field stars to calibrate the photometry, and are included in Figure 6 and Table 5. We have both an unfiltered and an \emph{R}-band image on day 14.9 after the assumed explosion date, and the magnitudes measured from these are consistent within the errors. 

In the \emph{UBgVIJHK} bands the epoch of maximum light cannot be determined. The uncertainty of the early \emph{R}-band data points makes it difficult to determine it precisely; in the \emph{r} band the uncertainty is smaller, but the first three data points are still consistent with each other. However, from the fits in Table 4, assuming similarity with SN 1998S and a rise time of 15 d, we can estimate that it happened on day 7 -- 8 after discovery. For SN 1998S the peak was only observed in the \emph{R} band. Therefore we adopt the epoch of the brightest \emph{R}-band data point, 7.2 days after discovery (14.9 d after explosion), as the maximum; this is also consistent with the brightest \emph{r}-band point. The \emph{i}-band peak is approximately on day 17 -- 19 after explosion.

\begin{table*}
\begin{minipage}{128mm}
\centering
\caption{The peak absolute \emph{B} magnitude of SN 2013fc compared to some other type II SNe. The distances used are the most recent values (from the Cepheid or Tully-Fisher method) reported on NED as of August 2015.} 
\begin{tabular}[t]{lcccc}
    \hline
	SN & $M_{B,max}$ & Peak \emph{B} ref. & $D$ (Mpc) & $D$ ref.\\
    \hline
    \hline
	2013fc & -20.46 $\pm$ 0.21 & this work & 83.2 & \citet{cat1} \\
	1998S & -19.43 $\pm$ 0.15 & \citet{1998S2} & 15.2 & \citet{dist98S} \\
	1979C & -20.32\footnote{No photometric errors were reported for SN 1979C.} & \citet{1979C1} & 17.2 & \citet{dist79C} \\
	2008fq & -18.34 $\pm$ 0.01\footnote{The extinction of SN 2008fq was determined through the EW of its Na {\sc i} D absorption, which has been shown to be unreliable \citep{naid}; it was likely brighter than reported here (see below).} & \citet{taddia13} & 35.6 & \citet{dist08fq} \\
    \hline
	mean II-L & -17.98 $\pm$ 0.34 & \citet{p15} &  &  \\
	mean II-P & -16.75 $\pm$ 0.37 & \citet{p15} &  &  \\
	mean IIn & -18.53 $\pm$ 0.32 & \citet{p15} &  &  \\
\hline
\end{tabular}
\end{minipage}
\end{table*}

Assuming the aforementioned total extinction value of $A_{V} = 2.9$ mag, we obtain a peak \emph{B}-band absolute magnitude of $-20.46 \pm$ 0.21 (from the first data point in this band, which was 1 d before the \emph{R}-band maximum); the error includes the uncertainties of the photometry, the distance and the extinction. This value is compared to the peak magnitudes of other type II SNe, including some 1998S-like SNe, in Table 8. 

The optical light curve shows four phases after the maximum (day 14.9): i) a decline phase until day $\sim$40; ii) a short plateau until day $\sim$65, which is not visible in \emph{B} or \emph{g}; iii) the following steeper decline until day $\sim$80; and iv) the tail phase with a linear decline from day $\sim$80 onward. SN 1998S had phases similar to these (see the lower panel of Figure 6), and the shapes of the optical light curves match remarkably well. The NIR light curves start on day 52 and show hints of the plateau and tail phases, especially in the \emph{J} band. In the NIR bands, the difference between SNe 2013fc and 1998S grows beyond the aforementioned 0.9 mag. These phases are described in more detail below. In Figure 7, we show a comparison between the absolute \emph{V}-band light curves of SN 2013fc and several other SNe. The phases described above are seen to some degree in other 1998S-like and type II-L SNe; but the light curves of the SNe in the lower panel, representing various different type II SNe, are quite different. For example, SN 2010jl \citep{2010jl} reached a comparable brightness but its light curve evolution was very slow.

In the first phase, our data show a linear decline in all optical bands. In the NIR bands we have no data from this phase. The rate of decline (here and afterwards calculated by least-squares fitting) is faster for the bluer wavelength bands: between 0.024 mag d$^{-1}$ in \emph{I} and 0.074 mag d$^{-1}$ in \emph{U}. In the \emph{U} band our data are sparse, and we have included the 51 d point in the least squares fit to determine the rate. Similar decline rates are seen in 1998S-like SNe \citep{1998S2} and type II-L SNe in general \citep[e.g.][]{1979C1, 80k}. The \emph{V} and \emph{R} magnitudes both decline $> 0.5$ mag from their peak values in this phase; thus, applying the definitions of \citet{faran2} and \citet{p9} respectively, SN 2013fc is consistent with type II-L.

The second phase contains first a short plateau, followed by a steeper drop in brightness in the third phase. This is seen in the \emph{VrRiI} bands. The \emph{V}-band decline is 0.021 mag d$^{-1}$ during the plateau and 0.082 mag d$^{-1}$ during the drop. The plateau ends around day 65 in these bands. Compared to type II-P SNe the drop from the plateau is not very pronounced, roughly 1.4 mag in \emph{V}. This phase is very similar in SN 1998S. Other 1998S-like events also show this phase to some degree; in fact, it has been suggested by \citet{p6} and \citet{13by} that most type II-L SNe, if followed long enough, would show this short plateau and subsequent faster decline (which is another sign of a continuum between types II-L and II-P); \citet{p6} find very few truly linear events in their sample. The \emph{B}-band light curve of neither SN 2013fc nor SN 1998S shows any sign of a plateau or a steeper drop. In the NIR, our data are not sufficient to see the beginning of the plateau, and in the \emph{HK} bands the decline until day 118 is quite linear. In the \emph{J} band, however, we do see a hint of a modest brightness drop of $\sim 0.5$ mag at roughly the same epoch as in \emph{I}. In the NIR, the brightness difference between SN 2013fc and SN 1998S begins to grow greater than the 0.9 magnitudes we see in the optical, although to a lesser extent in \emph{J}.

The last phase, beginning after the brightness drop, is linear in the \emph{BgVrRiIJ} bands; the decline rates are between 0.016 and 0.025 mag d$^{-1}$. The \emph{r}-band light curve seems to flatten off in this phase, but as the \emph{R}-band brightness declines at 0.019 mag d$^{-1}$, this is likely not the effect of a strong H\,$\alpha$ line but a systematic effect in the late-time \emph{r}-band data. \citet{1998S2} reported that the tail phase decline of SN 1998S was close to that expected from the decay of $^{56}$Co, but this only applies in the \emph{R} band, where the decline is 0.013 mag d$^{-1}$; in other bands the tail phase was similar to SN 2013fc with rates between 0.015 and 0.020 mag d$^{-1}$. A fast tail-phase decline like this also applies to SN 1979C, where the rates are over 0.02 mag d$^{-1}$ in the \emph{B} and \emph{V} bands. The light curve luminosity may be powered predominantly by CSM interaction at this stage instead of the $^{56}$Co decay, resulting in a faster decline than expected from a radioactive tail phase.

In the \emph{H} and \emph{K} bands, we instead see a drop in brightness after day 118 until about day 141, after which the light curve flattens off to some degree. The flattening was also seen in SN 1998S, where \emph{K}-band emission continued at roughly the same level from about day 100 to day 400 \citep{pozzo}, and \emph{H} also declined very slowly until then. SN 1979C also exhibited a very similar NIR evolution \citep{meikle}, with peak NIR magnitudes around $-20$ and a flattening late-time light curve. Detectable emission from SN 1998S continued in the NIR bands for more than a thousand days. This may also be the case with SN 2013fc, given the other similarities. A subtraction between our NIR template sets, corresponding to epochs of 406.9 and 740.8 d, does not show a detectable difference in the brightness of the coincident bright spot. This does not necessarily mean the SN has disappeared, but that its NIR brightness has stayed roughly constant between the sets. The NIR brightness difference between SN 1998S and SN 2013fc drops to about 0.9 magnitudes or even less in this phase; but if our NIR templates, in fact, still contain the SN, this would mean that the true NIR emission, especially in the tail phase, would be brighter than what we report here. However, SN 1998S declined $> 1$ mag in \emph{JHK} between days 405.5 and 698.2 \citep{pozzo}, so it is perhaps more likely that by day $\sim$ 400, the epoch of our first NIR templates, the coincident bright spot already completely dominated over SN 2013fc.

\begin{figure}
\centering
\includegraphics[width=\columnwidth]{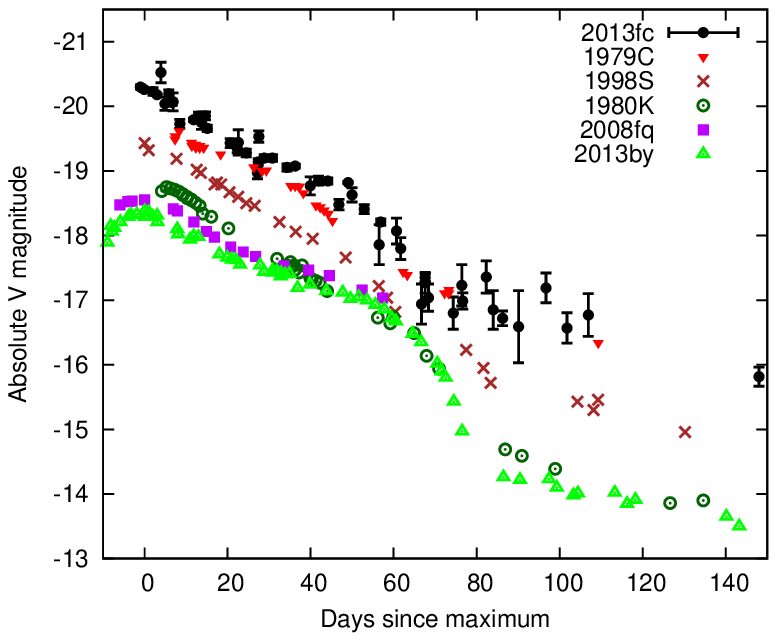}
\includegraphics[width=\columnwidth]{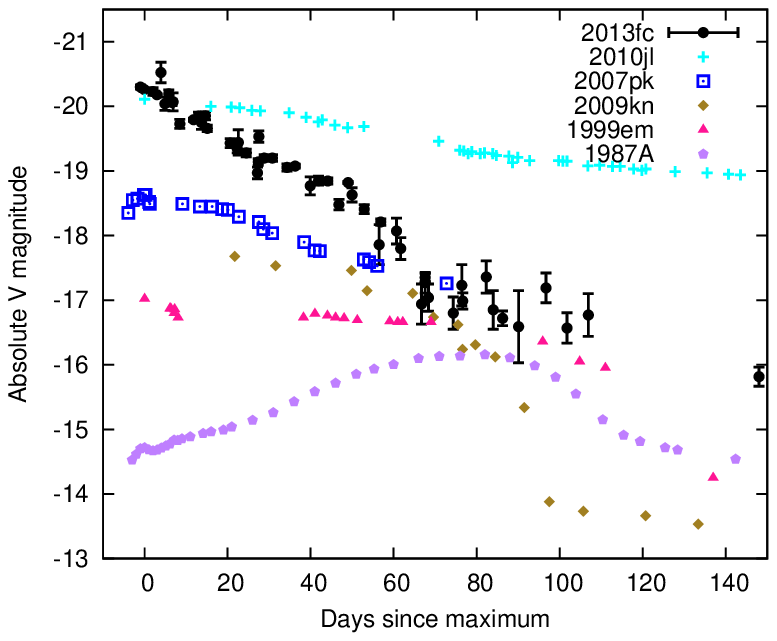}
\caption{Upper panel: the absolute \emph{V}-band light curve of SN 2013fc, compared to SNe 1979C \citep{1979C1}, 1998S \citep{1998S2}, 1980K \citep{80k, 80k2}, 2008fq \citep{taddia13} and 2013by \citep{13by}. All of these events show a II-L-like initial fast decline followed by a plateau phase and a drop after it. Lower panel: the absolute \emph{V}-band light curve of SN 2013fc, compared to other historical type II SNe with different light curves: SNe 2010jl \citep[IIn;][]{2010jl}, 2007pk \citep[intermediate II-P/II-L;][]{07pk2}; 2009kn \citep[IIn;][]{09kn}, 1999em \citep[II-P;][]{1999em} and 1987A \citep[II peculiar;][]{87a2}. In the case of SN 1987A, the maximum epoch refers to the first peak of the light curve, while for SN 1979C we use the \emph{B}-band maximum. All light curves are corrected for extinction.}
\end{figure}

\subsection{Colour evolution}

\begin{figure}
\centering
\includegraphics[width=\columnwidth]{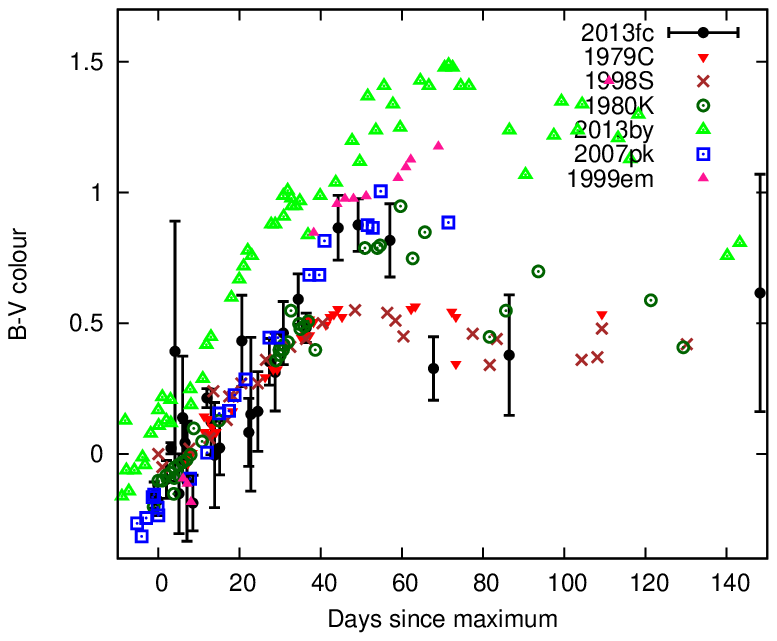}
\includegraphics[width=\columnwidth]{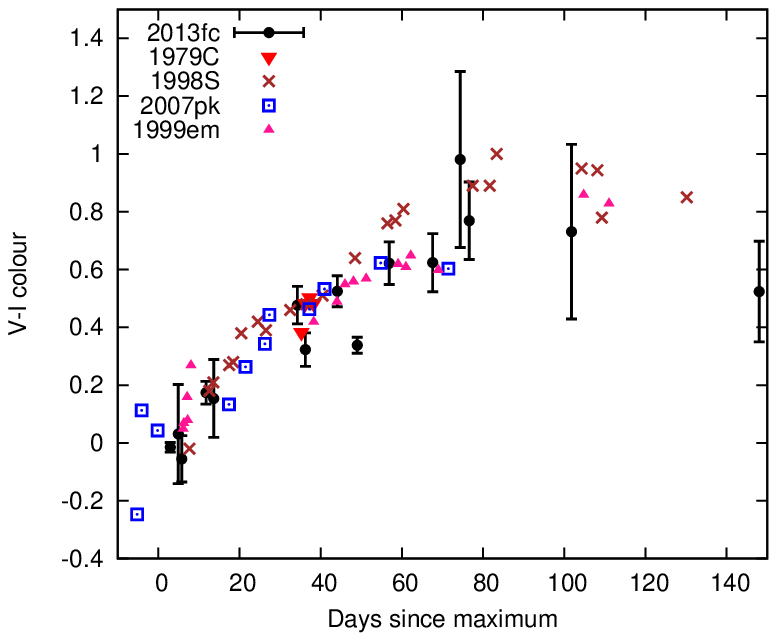}
\caption{The $B-V$ (upper panel) and $V-I$ (lower panel, if \emph{I}-band data was available) colour evolution of SN 2013fc, compared to similar SNe 1979C and 1998S and to other type II SNe with a plateau phase and a subsequent drop. All colours are corrected for extinction.}
\end{figure}

The $B-V$ and $V-I$ colour evolution of SN 2013fc compared to the similar SNe 1979C and 1998S, as well as other different type II SNe showing a plateau and a drop in their light curve, is shown in Figure 8. The LCOGT light curves seem to become quite noisy at $\gtrsim 20$ mag; these magnitudes are reached on day $\sim 55$ in \emph{B}. Therefore we have not included the LCOGT data points from day 55 onward in Figure 8. As mentioned previously, the extinction of SN 2008fq was estimated from the EW of its Na {\sc i} D line; thus SN 2008fq is not included in Figure 8, but the slope of its colour evolution is similar to the other 1998S-like events.

Until about day 25 after maximum, the $B-V$ evolution of all events in Figure 8 except SN 2013by looks similar. The $B-V$ colour of SN 2013fc changes from $\sim 0$ to $\sim 0.5$ mag between maximum and 37 days after maximum. At this point, the $B-V$ curves of SNe 1998S and 1979C flatten off at $\sim 0.5$ mag, while the $B-V$ of SNe 2013fc and 2007pk keeps following the prototypical type II-P SN 1999em. However, at about day 60 after maximum, the $B-V$ of SN 2013fc drops from the highest value of $\sim 0.85$ mag roughly to the same level as SNe 1979C and 1998S and stays consistent with them afterwards, within the errors, while SNe 1999em and 2013by keep turning redder. The epoch where SN 2013fc gets bluer again corresponds to the end of the short plateau phase, which can be seen in \emph{V} but not in \emph{B}. After this the colour stays roughly constant until the end of our optical light curve.

The $V-I$ evolution of all the events in the lower panel of Figure 8, on the other hand, looks fairly similar, although few \emph{I}-band data points of SN 1979C are available. The $V-I$ colour of SNe 2013fc, 1998S, 1999em and 2007pk is $\sim 0$ mag at maximum light and climbs to 0.6 -- 0.7 mag by day 60 after maximum. After this point, the $V-I$ colour curve flattens off as well, SN 2013fc with its $V-I$ value of $\sim 0.8$ mag being roughly consistent with SNe 1998S and 1999em; the light curve of SN 2007pk ends at the early phase of this flattening. Within the error bars, the flat $V-I$ curve continues to the end of our light curve observations (although after 140 days SN 2013fc seems slightly bluer than SNe 1998S and 1999em). As the data of SN 2013by reported by \citet{13by} were calibrated to the \emph{i} band instead of \emph{I}, it has not been included in the lower panel; however, its $V-i$ colour evolution is consistent with the $V-I$ evolution of the other events.

The colour evolution can be used as a consistency check for our extinction estimate. Assuming a lower $A_{V}$, such as 1 mag, would lower the brightness to a value more typical for type II-L. However, with $A_{V} = 1$ mag, the $B-V$ and $V-I$ colours of SN 2013fc would become $\sim0.6$ and $\sim0.7$ mag redder, respectively, making this event very different from the other type II SNe in Figure 8. As the NIR is only weakly affected by extinction, one would also see a very strong NIR excess. Thus we conclude that the high extinction of $A_{V} = 2.9$ mag we have determined earlier is indeed plausible and likely.

\subsection{Pseudo-bolometric luminosity}

\begin{figure}
\centering
\includegraphics[width=\columnwidth]{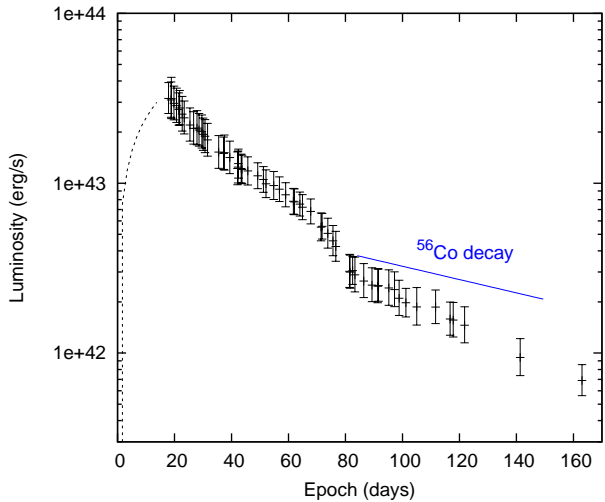}
\caption{The pseudo-bolometric light curve of SN 2013fc, calculated using the wavelength range of 3200 -- 23500 \AA. The dashed line represents the function $L = L_{\textrm{peak}} (t/t_{\textrm{peak}})^{2/3}$ used to estimate the pre-maximum luminosity.}
\end{figure}

Based on the photometry of SN 2013fc, and the luminosity distance to its host galaxy, we have estimated its pseudo-bolometric luminosity as follows. For each date, starting from our first \emph{U}-band data point, for which we have imaging in any band, we have linearly interpolated the magnitudes in the other bands to the specific date. The \emph{gri}-band data points end before the \emph{BVRIJHK} bands and have not been extrapolated to later times. For the NIR bands, for which we have no imaging data before day 52, we have linearly extrapolated the magnitudes for the epochs before that using the first six data points. This results in uncertainties in the flux in the beginning, but the contribution of the \emph{JHK} bands at these epochs is smaller than at later times ($\sim 15$ per cent on day 18 -- 22 vs. $\sim 40$ per cent around day 100). After de-reddening and applying the distance modulus, the absolute magnitudes were converted into fluxes using the zero-magnitude fluxes from \citet{mag1} and \citet{mag2}. We approximated the integral of the flux over the wavelength range of 3200 \AA~to 23500 \AA~using the trapezoidal rule, setting the fluxes at the end-points to zero. The end-points were chosen based on the limits of the \emph{U} and \emph{Ks} filters used. The resulting pseudo-bolometric light curve is plotted in Figure 9.

We then calculated the total radiated energy by integrating over the pseudo-bolometric light curve. The trapezoidal rule was used again, with the flux at the estimated epoch of explosion set to zero. The rise to peak was approximated with the function $L = L_{\textrm{peak}} (t/t_{\textrm{peak}})^{2/3} $, which was found to give a good fit to the pre-maximum \emph{R}-band light curve of SN 1998S from \citet{98s_early}. This was then integrated up to $t_{\textrm{peak}}$ to estimate the pre-maximum emitted energy. The resulting total radiated energy over the measured light curve (0 -- 163 days) is estimated at $1.13_{-0.06}^{+0.07} \times 10^{50}$ erg. The highest pseudo-bolometric luminosity values are $\sim 3.2 \pm 0.7 \times 10^{43}$ erg s$^{-1}$.

The pseudo-bolometric light curve, like in the case of SN 1998S, shows the same general phases as the optical light curves: linear decline, slower decline (not as clear as the short plateau in the \emph{VrRiI} light curves due to the influence of \emph{U}, \emph{g} and \emph{B}), drop and linear tail. The least-squares fit to the decline of the tail phase yields $0.0198 \pm 0.0007$ mag d$^{-1}$. Despite the tail phase likely being powered by CSM interaction instead of radioactivity (see Section 6), we can use the tail phase of the pseudo-bolometric light curve to estimate an upper limit for the $^{56}$Ni mass ejected by the SN. Using the same method we used to calculate the pseudo-bolometric luminosities for SN 2013fc, we did it for SN 1987A as well (taking the photometry from Burki et al. 1991). This was done to reduce systematic differences between the luminosities. The early and late tail-phase pseudo-bolometric luminosity of SN 1987A were then compared to those of SN 2013fc at the corresponding dates ($\sim$120 d, corresponding to the beginning of the tail phase in SN 1987A, and $\sim160$ d, corresponding to the end of our pseudo-bolometric light curve) to get the ratio of $^{56}$Ni masses. With a $^{56}$Ni mass of $0.069 M_{\odot}$ \citep{87a} for SN 1987A, we obtain estimates of $0.28_{-0.06}^{+0.08} M_{\odot}$ and $0.24_{-0.05}^{+0.06} M_{\odot}$ for SN 2013fc. Due to the fast decline, the later estimate gives a value closer to the actual $^{56}$Ni mass; thus we arrive at a limit of $\le0.30 M_{\odot}$.

\subsection{Blackbody fitting}

\begin{table*}
\centering
\begin{minipage}{174mm}
\caption{The parameters of the blackbody fits to the photometry of SN 2013fc -- temperature; radius; photospheric velocity, assuming a radius of zero the time of the explosion; and the blackbody luminosity of the hot component. The early-time values of $v_{phot,1}$ given in brackets correspond to a 5-day shift in the explosion date, which gives a better match to the temperature evolution of SN 1998S; later on the difference becomes small. The fits before day 75.6 only contain one component.} 
\begin{tabular}[t]{lcccccccc}
    \hline
        Epoch\footnote{Since explosion, in the rest frame of the SN.} & $T_{1}$ & $R_{1}$ & $v_{phot,1}$ & $L_{BB,1}$ & $T_{2}$ & $R_{2}$ & $v_{phot,2}$ & $L_{BB,2}$ \\
	(days) & ($10^{3}$ K) & ($10^{15}$ cm) & ($10^{3}$ km s$^{-1}$) & ($10^{42}$ erg s$^{-1}$) & ($10^{3}$ K) & ($10^{15}$ cm) & ($10^{3}$ km s$^{-1}$) & ($10^{42}$ erg s$^{-1}$)\\
    \hline
    \hline
	17.9 & 11.5 $\pm$ 0.9 & 2.07 $\pm$ 0.36 & 13.4 (10.5) $\pm$ 2.3 & 53.4 & - & - & - & - \\
	19.8 & 10.5 $\pm$ 0.9 & 2.25 $\pm$ 0.42 & 13.2 (10.5) $\pm$ 2.5 & 43.9 & - & - & - & - \\
	21.8 & 9.7 $\pm$ 0.7 & 2.39 $\pm$ 0.43 & 12.7 (10.3) $\pm$ 2.3 & 36.0 & - & - & - & - \\
	23.4 & 9.2 $\pm$ 0.6 & 2.52 $\pm$ 0.44 & 12.5 (10.3) $\pm$ 2.2 & 32.4 & - & - & - & - \\
	25.4 & 8.5 $\pm$ 0.5 & 2.70 $\pm$ 0.47 & 12.3 (10.3) $\pm$ 2.1 & 27.1 & - & - & - & - \\
	27.9 & 8.5 $\pm$ 0.6 & 2.64 $\pm$ 0.47 & 11.0 (9.3) $\pm$ 2.0 & 25.9 & - & - & - & - \\
	29.6 & 8.1 $\pm$ 0.6 & 2.71 $\pm$ 0.50 & 10.6 (9.1) $\pm$ 2.0 & 22.5 & - & - & - & - \\
	31.6 & 7.8 $\pm$ 0.6 & 2.80 $\pm$ 0.53 & 10.3 (8.9) $\pm$ 1.9 & 20.7 & - & - & - & - \\
	35.5 & 7.6 $\pm$ 0.5 & 2.70 $\pm$ 0.49 & 8.8 (7.8) $\pm$ 1.6 & 17.3 & - & - & - & - \\
	37.5 & 7.1 $\pm$ 0.5 & 2.95 $\pm$ 0.55 & 9.1 (8.0) $\pm$ 1.7 & 15.8 & - & - & - & - \\
	39.4 & 7.0 $\pm$ 0.4 & 3.00 $\pm$ 0.55 & 8.8 (7.8) $\pm$ 1.6 & 15.4 & - & - & - & - \\
	42.1 & 6.8 $\pm$ 0.4 & 2.99 $\pm$ 0.54 & 8.2 (7.3) $\pm$ 1.5 & 13.6 & - & - & - & - \\
	43.7 & 7.0 $\pm$ 0.4 & 2.76 $\pm$ 0.48 & 7.3 (6.6) $\pm$ 1.3 & 13.0 & - & - & - & - \\
	45.7 & 6.5 $\pm$ 0.4 & 3.08 $\pm$ 0.55 & 7.8 (7.0) $\pm$ 1.4 & 12.1 & - & - & - & - \\
	52.1 & 6.4 $\pm$ 0.3 & 2.85 $\pm$ 0.45 & 6.3 (5.8) $\pm$ 1.0 & 9.7 & - & - & - & - \\
	54.8 & 6.3 $\pm$ 0.3 & 2.85 $\pm$ 0.44 & 6.0 (5.5) $\pm$ 1.0 & 9.1 & - & - & - & - \\
	56.8 & 6.5 $\pm$ 0.2 & 2.75 $\pm$ 0.42 & 5.6 (5.2) $\pm$ 0.9 & 9.6 & - & - & - & - \\
	59.0 & 6.3 $\pm$ 0.2 & 2.81 $\pm$ 0.43 & 5.5 (5.1) $\pm$ 0.9 & 8.9 & - & - & - & - \\
	62.1 & 5.9 $\pm$ 0.2 & 2.99 $\pm$ 0.46 & 5.6 (5.2) $\pm$ 0.9 & 7.7 & - & - & - & - \\
	64.0 & 6.5 $\pm$ 0.3 & 2.58 $\pm$ 0.41 & 4.7 (4.3) $\pm$ 0.8 & 8.5 & - & - & - & - \\
	67.8 & 6.0 $\pm$ 0.2 & 2.73 $\pm$ 0.43 & 4.7 (4.3) $\pm$ 0.8 & 6.9 & - & - & - & - \\
	71.8 & 5.6 $\pm$ 0.2 & 2.86 $\pm$ 0.45 & 4.6 (4.3) $\pm$ 0.8 & 5.7 & - & - & - & - \\
     \hline
	75.6 & 5.9 $\pm$ 0.7 & 2.25 $\pm$ 0.59 & 3.4 $\pm$ 0.9 & 4.4 & 2.0 $\pm$ 0.5 & 6.14 $\pm$ 2.26 & 8.8 $\pm$ 3.2 & 0.5\\
	81.6 & 5.4 $\pm$ 0.8 & 2.01 $\pm$ 0.61 & 2.9 $\pm$ 0.9 & 2.4 & 2.0 $\pm$ 0.5 & 7.75 $\pm$ 2.06 & 10.4 $\pm$ 2.8 & 0.6\\
	86.4 & 6.7 $\pm$ 1.3 & 1.06 $\pm$ 0.38 & 1.4 $\pm$ 0.6 & 1.6 & 2.2 $\pm$ 0.3 & 6.76 $\pm$ 1.24 & 8.6 $\pm$ 1.6 & 1.1 \\
	89.3 & 6.2 $\pm$ 1.2 & 1.25 $\pm$ 0.45 & 1.6 $\pm$ 0.6 & 1.6 & 2.1 $\pm$ 0.4 & 7.31 $\pm$ 1.39 & 9.0 $\pm$ 1.7 & 0.9 \\
	91.6 & 6.4 $\pm$ 1.0 & 1.33 $\pm$ 0.38 & 1.7 $\pm$ 0.5 & 2.1 & 2.1 $\pm$ 0.3 & 8.84 $\pm$ 1.72 & 10.6 $\pm$ 2.1 & 0.9 \\
	95.2 & 7.4 $\pm$ 1.8 & 1.08 $\pm$ 0.43 & 1.3 $\pm$ 0.6 & 2.5 & 2.0 $\pm$ 0.4 & 7.83 $\pm$ 1.77 & 9.0 $\pm$ 2.0 & 0.8\\
	98.8 & 6.9 $\pm$ 1.7 & 1.09 $\pm$ 0.47 & 1.3 $\pm$ 0.6 & 1.9 & 2.1 $\pm$ 0.4 & 7.07 $\pm$ 1.56 & 7.9 $\pm$ 1.7 & 0.8 \\
	101.2 & 6.4 $\pm$ 0.9 & 1.19 $\pm$ 0.37 & 1.4 $\pm$ 0.5 & 1.7 & 2.1 $\pm$ 0.4 & 6.72 $\pm$ 1.53 & 7.3 $\pm$ 1.7 & 0.8 \\
	111.6 & 7.2 $\pm$ 1.5 & 1.00 $\pm$ 0.35 & 1.0 $\pm$ 0.4 & 1.9 & 2.1 $\pm$ 0.5 & 6.07 $\pm$ 1.44 & 6.0 $\pm$ 1.4 & 0.5\\
	116.7 & 6.7 $\pm$ 1.4 & 1.00 $\pm$ 0.37 & 1.0 $\pm$ 0.4 & 1.4 & 2.2 $\pm$ 0.5 & 5.42 $\pm$ 1.21 & 5.0 $\pm$ 1.2 & 0.5\\
	121.8 & 7.5 $\pm$ 1.9 & 0.78 $\pm$ 0.31 & 0.7 $\pm$ 0.3 & 1.4 & 2.2 $\pm$ 0.5 & 4.57 $\pm$ 1.21 & 4.2 $\pm$ 1.1 & 0.4\\
	141.3 & 6.8 $\pm$ 1.6 & 0.81 $\pm$ 0.34 & 0.7 $\pm$ 0.3 & 1.0 & 1.9 $\pm$ 1.2 & 2.25 $\pm$ 1.35 & 1.8 $\pm$ 1.1 & 0.1\\
	163.0 & 7.1 $\pm$ 1.2 & 0.62 $\pm$ 0.20 & 0.4 $\pm$ 0.2 & 0.7 & 2.4 $\pm$ 0.6 & 2.78 $\pm$ 0.72 & 1.9 $\pm$ 0.5 & 0.2 \\
\hline
\end{tabular}
\end{minipage}
\end{table*}

We have used the fluxes calculated from our photometry to fit blackbody functions. The first blackbody fit we include is from day 17.9, as this is when our \emph{U}-band light curve begins. The parameters of the blackbody fits are listed in Table 9, and those of the hot component are compared to those of SN 1998S \citep[][with the blackbody radii corrected for the more recent distance]{1998S2} in Figure 10. In each epoch, we have performed the fitting using the bands available, interpolating between epochs when necessary; no extrapolations have been done. Some redundant epochs have been omitted. The \emph{r} and \emph{R} bands, which include the H\,$\alpha$ emission line, have not been included after day 45, as the strength of the emission disrupts the shape of the continuum (see Section 5). Thus, for example, from day 17.9 to day 43.7, we use our \emph{UBgVrRiI} data, while between days 52.1 and 89.3 we have used \emph{BgViIJHK}. From day 75.6 onwards, one blackbody is no longer sufficient to provide good fits. Therefore, after this we have also fitted a secondary cool component after subtracting a fit to the optical bands. We tentatively attribute this component to dust emission, as the NIR light curves do not decline along with the optical ones. We show the fits and the fluxes in different bands on days 17.9, 75.6 and 116.7 in Figure 11. The blackbody luminosity on day 17.9, $\sim 5.3 \times 10^{43}$ erg s$^{-1}$, is $\sim 60$ per cent higher than the corresponding pseudo-bolometric luminosity due to the inclusion of the UV wavelengths; estimating the total radiated energy as we did for the pseudo-bolometric light curve, we get $\sim 1.4 \times 10^{50}$ erg. The difference between the blackbody and pseudo-bolometric luminosity decreases over time and becomes negligible at $\sim$ 40 days. 

\begin{figure}
\centering
\includegraphics[width=\columnwidth]{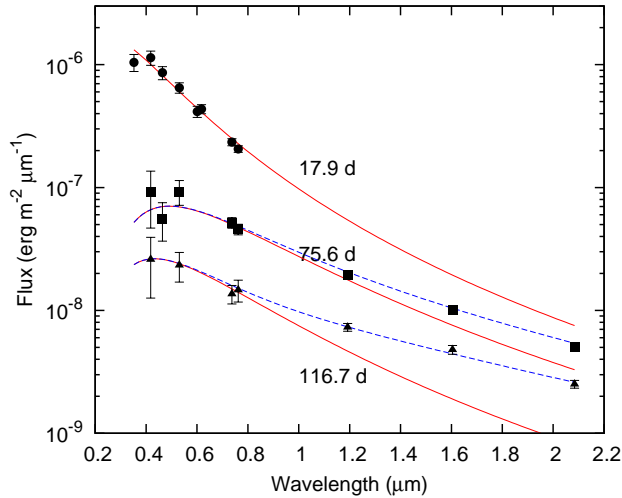}
\caption{The photometric fluxes of SN 2013fc in different bands on days 17.9, 75.6 and 117.8, compared with the blackbody fits in those epochs. The red lines correspond to the hot component only, while the blue dashed lines correspond to the sum of the two components.} 
\end{figure}

\begin{figure}
\centering
\includegraphics[width=\columnwidth]{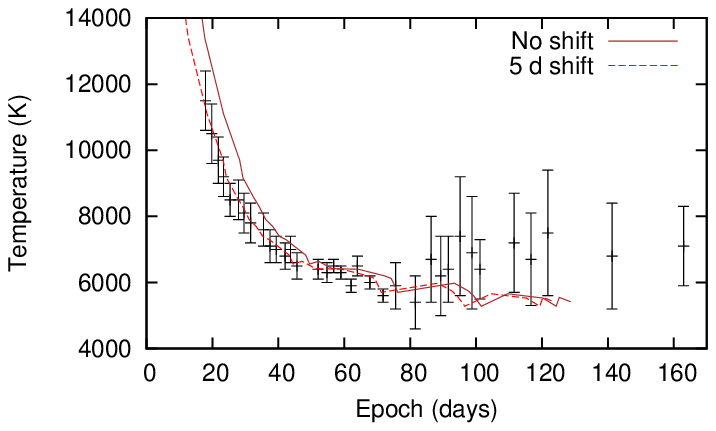}
\includegraphics[width=\columnwidth]{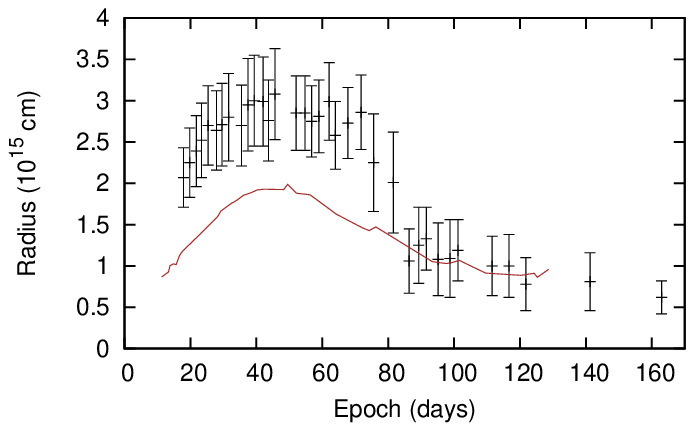}
\caption{The primary component of the blackbody fit of SN 2013fc (points) compared to SN 1998S (brown lines). The dashed red line in the upper panel illustrates the better fit to SN 1998S after a shift of the explosion date by 5 days.}
\end{figure}

The evolution of the blackbody parameters of SN 1998S \citep{1998S2} is fairly similar to the hot component of our fits. The difference in radius (by a factor of $\sim 1.5$) roughly corresponds to the brightness difference. The temperature of SN 2013fc is slightly lower than that of SN 1998S at early times, perhaps suggesting a required offset of a few days in our assumption of the explosion date -- the temperatures become consistent when one shifts the epochs of SN 2013fc by $\sim5$ days. The radii of the fits at late times are similar due to the SN 1998S fits only having one component, while the contribution of the cool component of SN 2013fc grows. The radius changes from about 2 $\times 10^{15}$ cm in the beginning to about $3 \times 10^{15}$ cm on day 30, after which it stays roughly constant until day $\sim 70$. The time when the photospheric radius stays constant approximately corresponds to the short plateau in the light curve. Meanwhile, the temperature of the hot component drops from $\sim$11500 K at day 18 to $\sim$7000 K around day 40, staying around 6000 -- 7000 K until the end of our observations. This suggests H recombination as a power source in this phase. From day 75 onwards, the blackbody radius starts to shrink and eventually, on day 163, has reached $\sim6 \times 10^{14}$ cm. The errors of the radius are large throughout, however, and toward the end, the parameters become physically less meaningful as the supernova becomes more nebular. The temperature of the second component seems to be roughly 2000 K, near the evaporation temperature of amorphous carbon, in all our two-component fits with no significant change from epoch to epoch. This suggests the existence of roughly 2000-K dust. The luminosity of this component, although extremely uncertain, seems to be between $\sim20$ and $\sim50$ per cent of the luminosity of the hot component before day 100, after which the value seems to decrease, although our \emph{HK} magnitudes may become unreliable at this point due to the possible continued presence of the SN in our template images. As the dust may only form a diluted blackbody or have a low covering factor due to clumpiness, the radii of the cool component are merely lower limits for the true outer extent of the dust.

\section{Spectroscopy}

In order to remove the contamination caused by the host galaxy (mainly from the H II/narrow-line region underlying the SN location), we have performed template subtractions for the imaging data. Similarly, we subtract the scaled 526 d WiFeS reference spectrum from the others, assuming that, at this point, the underlying region completely dominates over the SN spectrum\footnote{The PESSTO spectra available publicly in the ESO SAF are the observed spectra without the template subtraction.}. No spectral features of SN 2013fc can be seen in this spectrum, and the brightness of the SN had at this point faded below the detection threshold. However, the wavelength range of this spectrum is 3500 -- 7300 \AA. While reducing the EFOSC2 spectra of days 17, 27 and 51, we made extractions with background subtractions both next to the SN and using regions of background sky near the edges of the 2D spectrum. The difference between these was found to closely correspond to the shape of the 526 d spectrum, and the $>7300$ \AA~part was combined with the 526 d spectrum to create the final reference spectrum. 

We have scaled the final reference spectrum in such a way that the resulting subtraction provides the best match to the fluxes from the SN photometry after background subtraction. Inevitably, some residual of the narrow lines will remain; these have been removed from all the spectra in the rest of this paper through interpolation (this applies to the narrow H\,$\beta$, [O {\sc iii}] $\lambda\lambda4959,5007$, [N {\sc ii}] $\lambda\lambda6548,6583$, H\,$\alpha$ and [S {\sc ii}]$\lambda\lambda6717,6730$ lines). Additionally, the chip gaps at 4056 -- 4240 \AA~and 7315 -- 7490 \AA~in the RSS spectra have been similarly interpolated over (3985 -- 4165 \AA~and 7186 -- 7358 \AA~in the redshift-corrected spectra). All optical SN spectra from EFOSC2, RSS and WiFeS are presented in Figure 12, with the removed narrow lines marked by shaded regions. The spectra have been dereddened with $A_{V} = 2.9$ mag and corrected for redshift $z = 0.0179$, which, as mentioned in Section 3, was measured to be the average redshift of the narrow H\,$\alpha$ line from the underlying cluster that presumably contained the SN. We thus use this value as the rest frame of the SN. The evolution of the H\,$\alpha$ line profile can be seen in Figure 13.

\begin{figure*}
\centering
\begin{minipage}{150mm}
\includegraphics[width=\columnwidth]{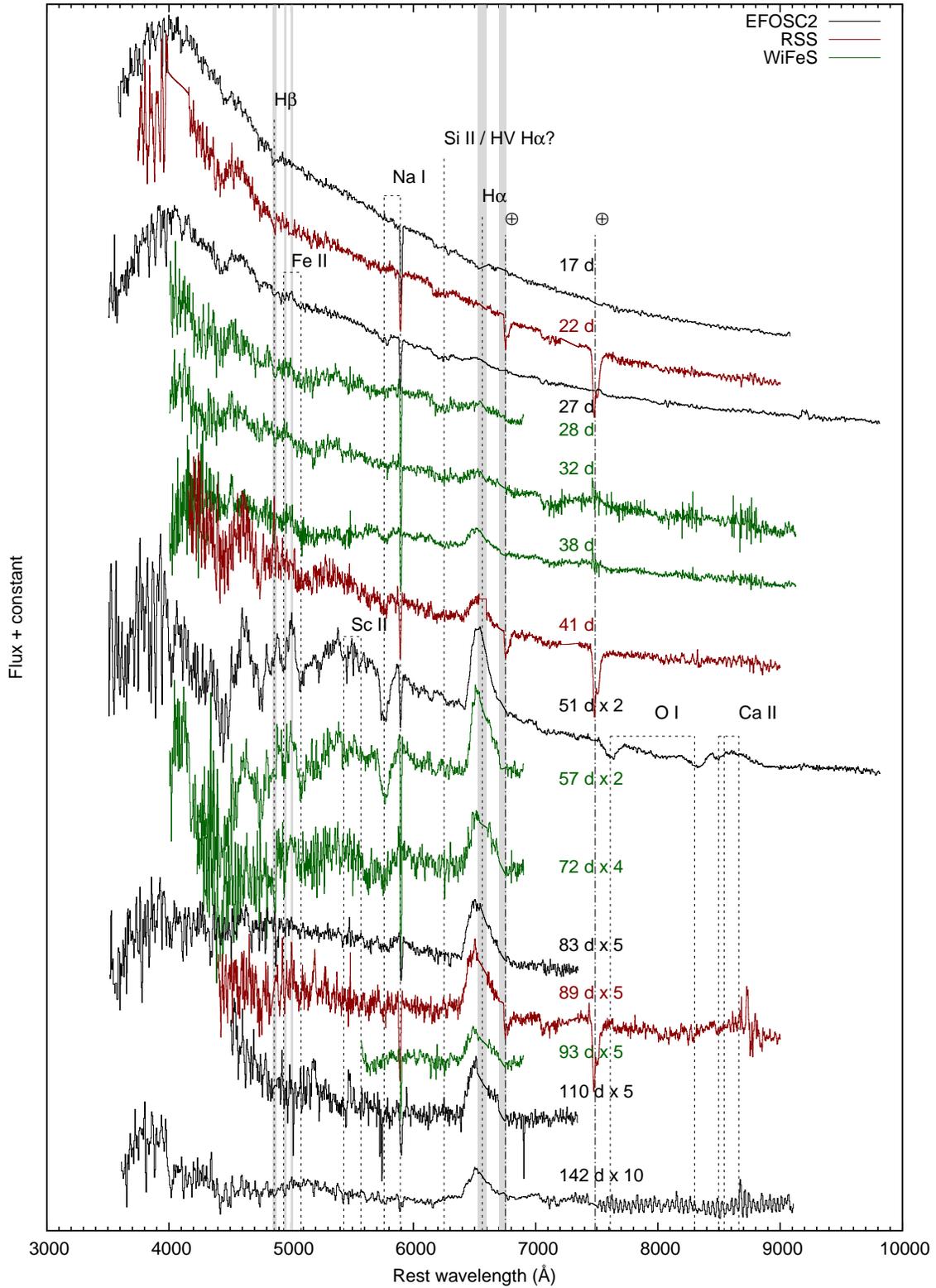}
\caption{The optical spectral evolution sequence of SN 2013fc. The very low signal-to-noise parts of especially the later spectra have been discarded for clarity. The spectra have been corrected for redshift and extinction. The shaded regions correspond to the removed narrow H\,$\beta$, [O {\sc iii}], H\,$\alpha$, [N {\sc ii}] and [S {\sc ii}] lines. The spectra after day 57 are vulnerable to small errors in the background removal due to the weakening SN continuum.}
\end{minipage}
\end{figure*}

\subsection{Early spectra (days 17 -- 41)}

The first spectrum in our set is from day 17. At this point the spectrum consists of a blue, almost featureless continuum. The only discernible lines, apart from the Na {\sc i} D absorption caused by the interstellar medium along the line of sight, are shallow H\,$\alpha$ and H\,$\beta$ absorption profiles. The H\,$\alpha$ absorption trough has a FWHM of $7600$ km s$^{-1}$ and a blueshift of $-2400$ km s$^{-1}$; the blue edge extends to $\sim-9000$ km s$^{-1}$, although the shallowness of the absorption compared to the noise makes precise determination of the edge practically impossible. The H\,$\beta$ absorption shows similar velocities. A weak, multi-component Si {\sc ii} $\lambda6355$ absorption feature can be seen, extending to a velocity of $\sim-10000$ km s$^{-1}$. \citet{07pk2} identify a similar absorption as high-velocity (HV) H\,$\alpha$ in the case of SN 2009dd, at a velocity of $\sim-14000$ km s$^{-1}$. Such a feature has been predicted by \citet{hvha} to appear at velocities as high as $\gtrsim10000$ km s$^{-1}$ in the photospheric phase of type II SNe. If this feature is attributed to H\,$\alpha$, it corresponds to velocities between $-19000$ and $-12000$ km s$^{-1}$, centred on $\sim-15500$ km s$^{-1}$; however, the lack of a clear HV H\,$\beta$ line supports the line being Si {\sc ii} rather than H\,$\alpha$.

The continuum then cools down, while the broad H\,$\alpha$ absorption first disappears by day 22 and then starts to turn into an emission line, with FWHM $6000$ km s$^{-1}$ on day 27. The Si {\sc ii} absorption is now more clearly visible. Other changes between these spectra include the appearance of broad absorption lines of Na {\sc i} D (possibly blended with He {\sc i} $\lambda5876$, as we also see possible He {\sc i} $\lambda4471$ absorption in the 17 d, 22 d and 27 d spectra, extending to $\sim -7500$ km s$^{-1}$; however, we can identify no other He {\sc i} lines) with a blueshift of $\sim-5500$ km s$^{-1}$ and a FWHM of $\sim4000$ km s$^{-1}$. Possible blueshifted Fe {\sc ii} absorption lines appear around 5000.

Between days 27 and 32, the spectrum stays similar apart from the cooling temperature, with practically the only other change being the broadening of the H\,$\alpha$ emission to $\sim8000$ -- 9000 km s$^{-1}$ as the absorption component keeps weakening (although the WiFeS spectra of days 28 and 32 are rather noisy). However, toward day 41 there is noticeable spectral evolution. On day 38, the H\,$\alpha$ emission and the broad Na {\sc i} D absorption have strengthened considerably. The H\,$\alpha$ profile has taken an asymmetric shape, with a steeper fall to continuum level on the blue side. The 41 d spectrum shows similar lines; the Fe {\sc ii} absorption lines are strengthening at this point. 

\subsection{Later spectra (days 51 -- 142)}

\begin{figure}
\centering
\includegraphics[width=\columnwidth]{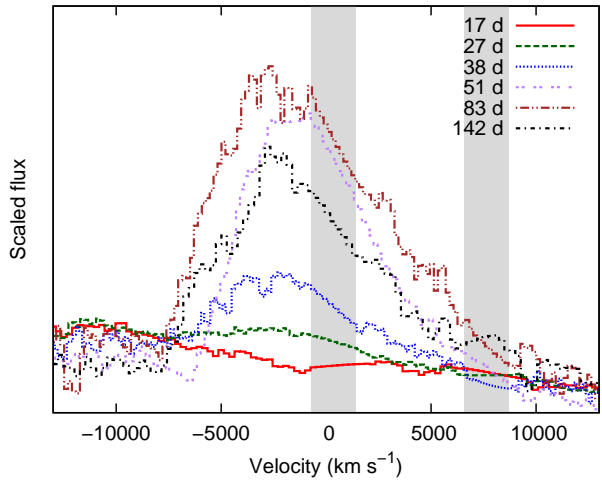}
\caption{The evolution of the velocity profile of the H\,$\alpha$ line, showing the change from a shallow absorption to strong asymmetric emission. The continua have been scaled to match on the red side of the line. The shaded region corresponds to the removed narrow H\,$\alpha$, [N {\sc ii}] and [S {\sc ii}] lines.}
\end{figure}

\begin{figure}
\centering
\includegraphics[width=\columnwidth]{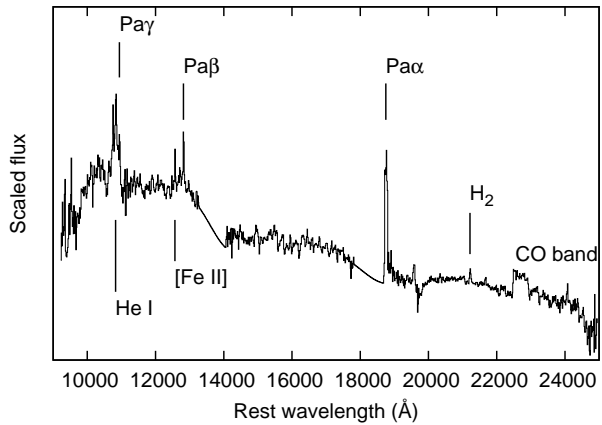}
\caption{The 82 day NIR spectrum of SN 2013fc. The spectrum is corrected for redshift and extinction, and telluric bands have been removed. The host galaxy background has not been removed.}
\end{figure}

The broad features have evolved to their full extent on day 51. The appearance of the forest of Fe {\sc ii} lines is by now clear; broad O {\sc i} $\lambda\lambda7774,8446$ absorption is visible at this point; the broad Na {\sc i} D absorption line is at its strongest and begins to also show an emission component creating a P Cygni shape. The blue edge of the absorption extends to $\sim -9000$ km s$^{-1}$. The H\,$\alpha$ emission is also at its strongest here, with an EW of $\sim200$ \AA; however, it shows evolution in other ways as well. The blue wing of the line is eroded by a possible new absorption component, which is also visible as a small highly blueshifted trough. This changes the FWHM of the emission to 7000 km s$^{-1}$ and moves the peak to a blueshift of 1100 km s$^{-1}$. The blue edge of the line is at $-6800$ km s$^{-1}$ (the trough). The red wing extends at least to the removed [S {\sc ii}] host galaxy line, which makes it difficult to accurately place the continuum-level edge; however, it appears to be roughly at $+8000$ km s$^{-1}$. The Si {\sc ii} absorption has faded to become indistinguishable from noise.

The weak absorption in the blue wing of the H\,$\alpha$ profile disappears by day 57, and the peak shifts back toward the red, to $-2300$ km s$^{-1}$. The line also broadens back to a FWHM of $\sim$8000 km s$^{-1}$, after which the shape hardly changes at all. Scandium absorption lines appear between the iron lines and the broad Na {\sc i} D feature. 

Between days 51 and 72, the broad Na {\sc i} D absorption weakens, but not much evolution can be seen in the Fe {\sc ii} lines; however, by day 83 the Fe {\sc ii} lines have disappeared and are not visible in our later spectra. On day 89, noisy Ca {\sc ii} emission is still visible, but it is undetected on day 142. The broad Na {\sc i} D absorption is also still tentatively detected on day 89, but not in the later spectra. In fact, our spectra from days 93, 110 and 142 show practically no clear features apart from the broad H\,$\alpha$ and the narrow Na {\sc i} D absorption. In any case, the H\,$\alpha$ profile still keeps its general shape until our last spectrum, with a FWHM of $\sim$8000 -- 9000 km s$^{-1}$ and the same asymmetric wings as before; the shape does show a hint of the profile becoming more `boxy' on day 142, but this is difficult to say for certain. Its EW also seems to stay between 150 and 200 \AA~from day 51 until the end of our observations, but this is also tentative due to the uncertain background correction of the late-time spectra.

The day 82 NIR spectrum is shown in Figure 14. The only feature that we detect clearly in this spectrum, apart from narrow H {\sc i}, H$_{2}$ 2.12 $\mu$m, He {\sc i} and [Fe {\sc ii}]$\lambda12570$ emission from the host galaxy, is a noisy He {\sc i} $\lambda10830$ feature with a broad P Cygni profile, which is consistent with the shape of the broad Na {\sc i} D in the optical spectra. The narrow Pa\,$\gamma$ and He {\sc i} lines from the host galaxy are overlaid on the broad He {\sc i} from the SN. The blue edge of the absorption extends to $\sim$ -9000 km s$^{-1}$, consistently with e.g. Na {\sc i} D in the optical spectra. There seems to be a broad Pa\,$\beta$ component with a FWHM between 8000 and 10000 km s$^{-1}$, but due to the noise this feature is unclear. The Pa\,$\alpha$ line looks like it may also have a broad component, but it falls on the wavelength region with very low atmospheric transmission between the \emph{H} and \emph{Ks} bands. A blueshifted CO first-overtone band can be seen as well, extending to $\sim$ -7500 km s$^{-1}$, roughly consistent with the H\,$\alpha$ and Na {\sc i} D.

\subsection{Comparison to similar events}

\begin{figure*}
\centering
\begin{minipage}{150mm}
\includegraphics[width=\columnwidth]{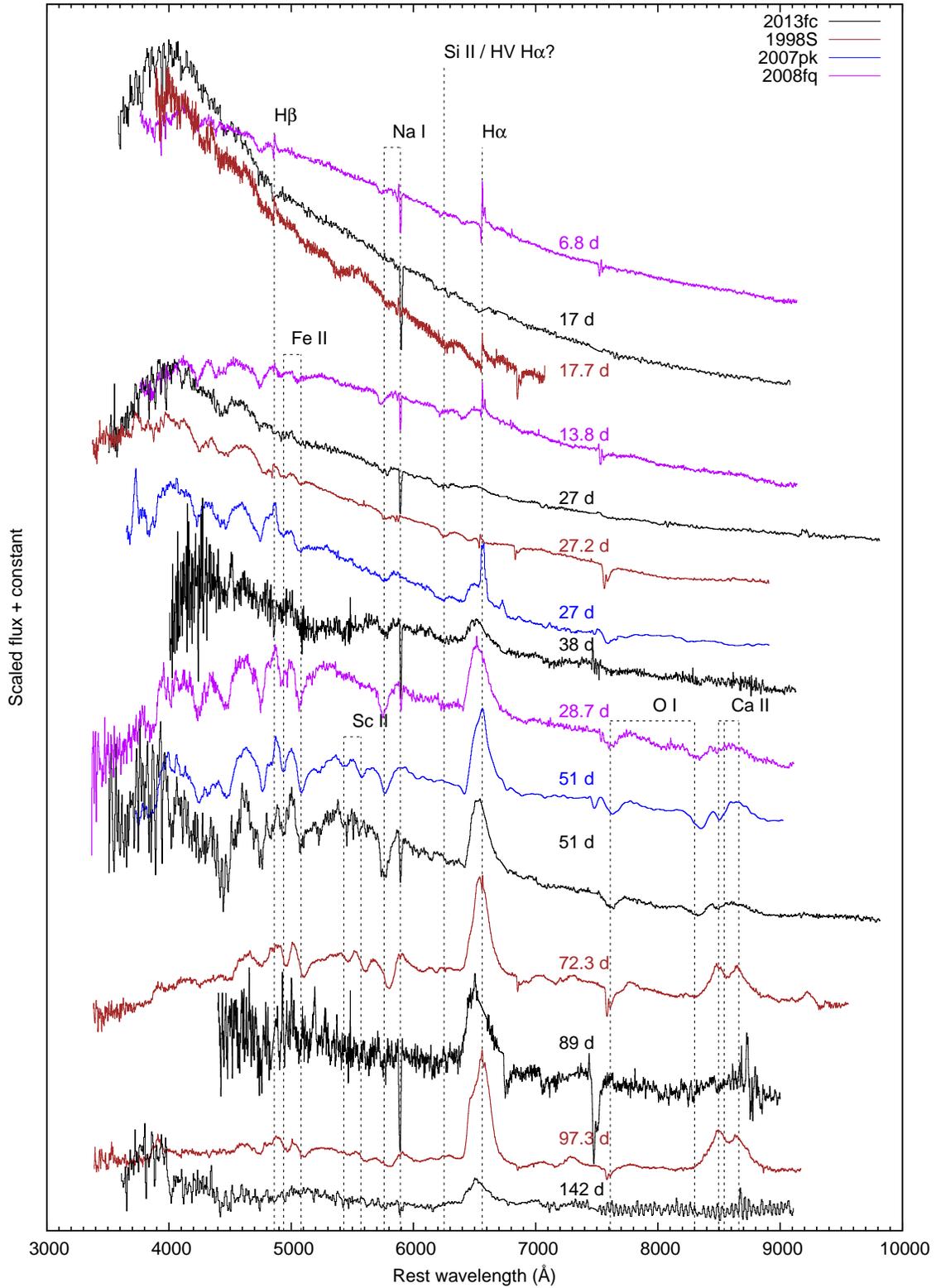}
\caption{Key spectra of SN 2013fc (black) compared to those of SN 1998S (brown), SN 2007pk (blue) and SN 2008fq (purple) at corresponding epochs.}
\end{minipage}
\end{figure*}

In Figure 15 we show a comparison of spectra of SN 2013fc and of SNe 1998S, 2007pk and 2008fq at roughly corresponding epochs, from \citet{1998S3}, \citet{07pk2} and \citet{taddia13} respectively, and obtained from WISeREP. Spectra of SNe 1979C, when mentioned, are from \citet{1979C5}. No SN 1979C spectra were found to be publicly available in ASCII form, and thus we do not include them in Figure 15.

Our first spectrum differs in some ways from SN 1998S on day 17.7. The Balmer lines in the SN 1998S spectrum show P Cygni profiles with a narrow emission and broader absorption, but this part of the SN 2013fc spectrum is dominated by the residuals of the narrow host galaxy lines. The closest SN 2008fq spectrum is from 6.8 d after discovery, corresponding to $<\sim16$ d after shock breakout according to \citet{taddia13}; the continuum of SN 2008fq looks redder (and continues to do so in later epochs), but as previously mentioned, its extinction may be unreliable. SN 2008fq exhibits broad, blueshifted absorption profiles of H and Na, with contamination from the host galaxy disrupting the narrow P Cygni from the SN. The day 27 spectrum of SN 2013fc is reminiscent of SN 1998S on day 27.2. The H\,$\alpha$ line profiles of all three events are changing from absorption to pure emission, but as the Balmer line absorption in SN 2013fc was weaker to begin with, it is the only one of the three at this point that only shows the H\,$\alpha$ emission without clear absorption. The early spectra of SN 1979C, however, show the broad but weak H\,$\alpha$ emission and an appearing Na {\sc i} absorption. 

The spectra of SNe 2007pk and 2008fq (apart from the narrow H\,$\alpha$ component in SN 2007pk, which is contamination from the host galaxy) evolve similarly to SN 2013fc but faster. The day 27 spectrum of SN 2007pk and the day 28.7 spectrum of SN 2008fq ($<\sim38$ days after shock breakout) exhibit clearer Fe {\sc ii} absorption. SN 2008fq also shows O {\sc i} absorption; its H\,$\alpha$ emission is somewhat stronger; and emission of the Ca {\sc ii} NIR triplet has appeared.

Our day 51 spectrum shows remarkable similarity to the SN 2008fq day 28.7 spectrum and the SN 2007pk day 51 spectrum. The shapes of the Na {\sc i} D P Cygni profile and O {\sc i} absorption lines are roughly equal to those of SNe 2008fq and 2007pk. SN 1979C on days 43 -- 72 also provides a good match, although no spectral information redwards from 7000 \AA~is available; its H\,$\alpha$ profile attains an asymmetric shape with a $-8500$ km s$^{-1}$ velocity at the blue edge. The H\,$\alpha$ profile of SN 2008fq also shows a similar shape, though with a small absorption trough at $-7500$ km s$^{-1}$, and a red wing extending to $\sim+$7500 km s$^{-1}$. The H\,$\alpha$ profile of SN 2007pk on day 51 is very similar to these SNe as well, and the Ca {\sc ii} NIR triplet is of similar shape and strength in the spectra of SNe 2013fc, 2008fq and 2007pk.

At later times, the Ca {\sc ii} NIR triplet becomes more prominent in SNe 1998S and 2008fq, but while it is visible on day 89 in SN 2013fc, it stays considerably weaker than in SN 1998S. Apart from the Sc lines and the strength of the Ca {\sc ii} emission, the day 72.3 spectrum of SN 1998S provides a good match to all SN 2013fc spectra between days 51 and 72. The FWHM of the H\,$\alpha$ profile is $\sim$7000 km s$^{-1}$, without a clear absorption trough on the blue side, the blue and red wings extending to $\sim-8000$ and $+9000$ km s$^{-1}$ respectively. Conversely, the absorption at the blue edge of the line never disappears from the spectra of SN 2007pk, instead getting stronger toward the later epochs. The evolution of SN 2013fc from day 83 onwards seems quite different from what happens to SN 1998S at similar epochs, but is probably due to the SN being fainter and the host galaxy contribution to the spectra being relatively stronger at later times, resulting in very noisy subtracted spectra that are susceptible to small changes in the scaling of the reference spectrum. Thus the only features we continue to see are the strong emission features, i.e. H\,$\alpha$ and the aforementioned Ca {\sc ii} on day 89. The shape of the H\,$\alpha$ profile of SN 1998S became more asymmetric and `boxy' between days 72 and 140 \citep{1998S5}; a hint of this evolution can be seen in SN 2013fc on day 142, but both later and weaker than in the case of SN 1998S.

\section{Discussion}

The light curve of SN 2013fc exhibits a linear fast decline followed by phases resembling the plateau and tail phase observed in II-P SNe. The rate of the linear decline is typical for historical type II-L SNe \citep[e.g. 1980K;][]{80k}, while the flattening phase is short and the brightness drop from it modest compared to the plateaus and post-photospheric-phase drops of type II-P SNe \citep[e.g. SN 1999em;][]{1999em}. However, it is abnormally bright, over two magnitudes brighter ($B_{max} = -20.46 \pm 0.21$) than the average for type II-L \citep[$-17.98 \pm 0.34$ mag;][]{p15} and over 3.5 magnitudes brighter than the average for type II-P ($-16.75 \pm 0.37$ mag). The spectra exhibit a nearly featureless, blue continuum that lasts $> 20$ days after the explosion. The H\,$\alpha$ profile (see Figure 13) does not show a clear P Cygni absorption, and consistently extends to velocities of $\sim \pm8000$ km s$^{-1}$. The profile is also consistently asymmetric, with the blue wing showing a steeper fall to continuum level. The narrow H\,$\alpha$ emission as a part of a multi-component profile seen in type IIn SNe is not visible, however.

These properties are similar to the 1998S-like SNe we have used for comparison \citep[e.g.][]{1979C1, 1979C5, 1998S2, 1998S3, taddia}. SN 2013fc is the brightest of these events, being $\sim$ 0.9 mag brighter than SN 1998S in \emph{BVRI} and mostly $\sim$0.6 mag brighter than 1979C in \emph{BV}. SNe 2013fc, 1979C and 1998S all show a flattening late-time NIR light curve; SNe 1979C and 2008fq reach NIR brightnesses comparable to SN 2013fc. The narrow Balmer lines were detected in SN 1998S, but only in the early phases; in the case of SNe 1979C and 2008fq, a narrow P Cygni H\,$\alpha$ profile can be seen in the early spectra, although it is very weak in the case of SN 1979C. As no pre-maximum spectra of SN 2013fc exist, and as the host galaxy Balmer lines are strong, we cannot exclude the early presence of narrow lines. This is speculative, however, and SN 2013fc should be considered a type II-L event. CSM interaction in itself does not necessarily require a type IIn classification either, as e.g. \citet{13by} has suggested that many type II-L events may involve some interaction.

The total radiated energy of SN 2013fc, $1.13_{-0.06}^{+0.07} \times 10^{50}$ erg, is $\sim10$ times higher than what is radiated by normal type II-P SNe over a similar period \citep[e.g.][]{energy1}. However, it is still a factor of a few to 10 lower than observed in the most luminous slowly declining type IIn SNe such as SN 1988Z \citep{1988z} and SN 2010jl \citep{2010jl}, and over 10 times lower than observed in superluminous type II events like SN 2006gy \citep[][]{06gy2, 06gy3, energy2} -- for an examination of superluminous SNe in general, see \citet{slsn}. The slowly-declining type IIn SNe can have peak magnitudes similar to SN 2013fc \citep[e.g.][]{kiewe,taddia}. However, their high total radiated energy is caused by strong interaction of the SN ejecta with a massive CSM, probably ejected in eruptive mass loss events.

SN 1998S, and to some degree SN 1979C, provide the basis for the progenitor discussion of SN 2013fc, by virtue of their similarity to SN 2013fc and being well-studied events. SN 1998S was interpreted as the explosion of a massive RSG based on its late-time luminosity \citep{mauerhan}, wind properties and CO emission \citep[][who also consider the possibility of pre-explosion blue supergiant evolution]{1998S3}. Mass loss was in the form of a strong RSG wind, possibly caused by pulsational instability \citep{mauerhan}. \citet{shivvers} fitted a radiative-transfer model to a very early (a few days after core collapse) high-resolution spectrum and found it consistent with a yellow hypergiant (YHG), an extreme RSG or an LBV progenitor (although CNO enrichment in the wind was found to be weaker than for Milky Way LBVs). Interaction of the SN ejecta with the inner dense CSM caused the early blue featureless continuum and a narrow H\,$\alpha$ emission profile with Lorentzian wings \citep{1998S3}. The material at the photosphere expanded at $\sim 9000$ km s$^{-1}$ \citep{1998S2}, and shock-deposited energy in the ejecta was released diffusively while the UV/X-ray emission from the flash and subsequent interaction was reprocessed by the CSM. A strong broad H\,$\alpha$ emission later appeared due to the ejecta interacting with outer (possibly asymmetric) CSM, with a possible contribution from a cool dense shell (CDS) at the ejecta-CSM interface \citep{1998S3, 1998S4, 1998S5}; such a shell is likely to form when the ejecta interact with CSM \citep{cds}. The asymmetric, blueshifted profile was caused by the receding part of the ejecta being hidden from sight, while the lack of a P Cygni absorption was due to the relatively low H envelope mass. CSM interaction continued to very late times \citep{1998S5}, creating a multi-peaked H\,$\alpha$ profile with high velocities even at day $\sim 1000$. A mass loss rate of $\sim2 \times 10^{-4} M_{\odot}$ yr$^{-1}$ was estimated by \citet{1998S6} using late-time radio observations. An infrared excess was caused by reradiation of UV and optical emission by dust in the CSM \citep[infrared echo;][]{1998S2}. The dust was argued to be pre-existing in the wind due to the lack of fast-moving metals in the early spectra and the required high velocity to get the dust-forming material to the observed distance, although further dust condensation in the CDS may also have occurred later \citep{pozzo, 1998S5}.

SN 1979C also exhibited similar spectral and photometric evolution \citep[e.g.][]{1979C1, 1979C5}. Strong narrow Balmer emission was not observed; only a weak narrow H\,$\alpha$ P Cygni profile can be seen in the early spectra superimposed on a weak broad H\,$\alpha$ emission. \citet{1979C2} proposed an RSG progenitor in a superwind stage, with a low remaining H envelope mass. It was also proposed that an even higher luminosity might be produced through a stronger mass loss \citep[this was invoked to explain the superluminous SN 2008es by][]{p13}. \citet{1979C7} also proposed an RSG progenitor, and found evidence of narrow UV lines originating in the CSM. The origin of broad H lines in the UV was proposed to be a thin shell close to the photosphere, heated and ionized by radiation from the shock. \citet{1979C6} proposed that the H\,$\alpha$ line should also originate close to the photosphere. The evolution of the NIR emission was quite similar to SN 1998S \citep{meikle}, with a long-lasting flat tail. Long-lasting CSM interaction was observed in the case of SN 1979C as well, in the form of radio emission \citep[e.g.][]{1979C4} and strong, broad very late-time optical line emission \citep{1979C8}. A mass-loss rate of $\sim 10^{-4} M_{\odot}$ yr$^{-1}$ over the last $\sim 10^{4}$ yr was derived from the radio data by \citet{1979C9} and \citet{1979C4}.

The observed evolution of SN 2013fc was similar to these events. The high peak luminosity, especially that of SNe 2013fc and 1979C, could be powered by early ejecta-CSM interaction or the ionization of a thick shell of CSM by the X-ray/UV flash. As mentioned before, the very early narrow lines are not detected, possibly due to a lack of early spectra; however, the presence of a blue featureless continuum as late as three weeks after the explosion still argues for early CSM interaction. As we show below, we see signs of dust condensing in the CDS. The broad H\,$\alpha$ profile is similar to SNe 1998S and 1979C; we also attribute its shape to opacity effects in the ejecta, possibly with electron scattering contributing to the red wing. The broad H\,$\alpha$ emission appears gradually, and a similar feature has been seen in multiple 1998S-like SNe. While the multi-layered CSM proposed for SN 1998S can be invoked to explain a singular or extremely rare SN, a nearly identical CSM structure (and thus mass-loss history) may be too much of a coincidence to be common to multiple SNe. A simple $\rho \propto r^{-2}$ CSM density profile may be adequate, as was assumed for SN 1979C by e.g. \citet{1979C4} and argued for SN 1998S by \citet{1998S5}. 

At early times, the narrow H\,$\alpha$ line would originate in the photoionized, unshocked CSM outside the ejecta. The blue featureless continuum would largely originate from the CDS, as suggested by \citet{1998S3}. As the ejecta expand, the narrow component would then weaken; the narrow P Cygni profile seen in SN 1998S could be rendered unobservable by the host galaxy contamination. The broad absorption seen in the earliest spectrum would originate from the ejecta close to the photosphere. The gradual appearance of the broad H\,$\alpha$ emission may be due to the photosphere receding through the ejecta as it becomes optically thinner, exposing the line-emitting region, until finally the entire region would lie outside the photosphere. The gradual appearance of the emission in SN 1998S may simply have been missed, while SN 1979C does show signs of this. At the same time the metal absorption lines (Na, Fe, Sc) appear as we begin to see the material deeper inside the ejecta. The optical light curves of SNe 2013fc and 1998S (and to some degree, other 1998S-like events) exhibit a very similar, simultaneous short plateau phase roughly at the same time as the broad H\,$\alpha$ emission appears. The plateau of SN 1998S was argued by \citet{1998S2} to be analogous to the recombination wave in a type II-P SN, which would place SN 1998S on the II-P/II-L continuum \citep[e.g.][]{p6}; the same likely applies to SN 2013fc. The blackbody temperature during the plateau phase in both SN 2013fc and SN 1998S is consistently over 6000 K, even over 7000 K when the phase begins; however, temperatures like this were also seen at similar times in e.g. the type II-P SN 2012aw \citep{2012aw}.

The luminosity of the H\,$\alpha$ emission, $\sim2.4 \times 10^{41}$ erg s$^{-1}$ on day 50 and $\sim5 \times 10^{40}$ erg s$^{-1}$ in the tail phase at ~100 days, is too high to be powered solely by the X-ray luminosity of the reverse shock. As for SN 2008es \citep{p13} we estimate \citep[following][]{cds} that the mass loss rate of the wind should be in excess of $10^{-3} M_{\odot}$ yr$^{-1}$ for this to work. Such a wind density would make the supernova look like a bona fide Type IIn, which is not the case. However, some powering of the H-alpha could be due to extreme UV photons and soft X-rays produced by inverse Compton scattering behind the forward circumstellar shock. As demonstrated by \citet{1979C9} for SNe 1979C and 1980K, such emission could easily dominate the ionizing radiation for the first months, with a luminosity in excess of a few times $10^{41}$ erg s$^{-1}$ after a few weeks, for a mass loss rate of $1.5 \times 10^{-4}$ $v/(\textrm{10 km s}^{-1})$ $M_{\odot}$ yr$^{-1}$. At late times, even though inverse Compton scattering may contribute, we find it likely that much of the H-alpha emission is due to release of stored shock energy from when the supernova shock traversed an extended envelope.

Some bright type II-L SNe, including SNe 1998S and 1979C, exhibit a late decline faster than the 0.98 mag (100 d)$^{-1}$ expected from a radioactive tail powered by $^{56}$Co decay. The reason for this may be inefficient $\gamma$-ray trapping in the SN ejecta. Such a scenario was suggested by \citet{p6}, who found that type II SNe with peak \emph{V} brightness $\lesssim -17$ mag tend to have \emph{V}-band tail-phase decline rates that are too steep for full $\gamma$-ray trapping. However, the tail phase may in fact be powered by continuing emission from the CSM interaction instead of or in addition to $^{56}$Ni \citep{1979C2}. In such a case, the $^{56}$Ni-powered luminosity could be significantly below the total luminosity. The $^{56}$Ni-powered tail of SN 1987A was about four times fainter than the late linearly declining phase of SN 2013fc at an epoch corresponding to the start of the tail in SN 1987A, despite the likely similar mass of the progenitor (see below); while the tail phase of SN 1979C, for which the presence of ejecta-CSM interaction has been invoked, was roughly as bright as that of SN 2013fc. Figure 18 of \citet{09md} shows a correlation in progenitor luminosity vs. apparent $^{56}$Ni mass, albeit with considerable scatter. Thus we tentatively favour the CSM interaction scenario as the explanation of the late-time decline rate (with perhaps a contribution from diffusion of stored energy produced by the supernova shock when it traversed the extended envelope), in which case the $^{56}$Ni mass estimate should in fact be considered an upper limit. Comparing the tail-phase luminosity of SN 2013fc against that of 1987A, we have thus arrived at a $^{56}$Ni mass limit of $\le0.30 M_{\odot}$. For SN 1998S, \citet{1998S2} estimated a $^{56}$Ni mass of $0.15 \pm 0.05 M_{\odot}$, but if this scenario is correct this value should also be taken as an upper limit of $\le0.20 M_{\odot}$. 

The presence of hot dust close to the evaporation temperature of amorphous carbon (2000 K) is suggested by our two-component blackbody fits. The dust radii are larger than those of the hot component, but no clear indication of increasing optical extinction is observed. Thus the covering factor and/or optical thickness are presumably low, possibly due to a clumpy structure. As in the case of SN 1998S, we see no signature of fast-moving metals in our early spectra. However, the radii of the cool blackbody component are far smaller than the dust evaporation radius we estimate for SN 2013fc. An evaporation radius of $5 \times 10^{16}$ cm can be calculated using Eq. 4 of \citet{radius2} assuming carbon-rich dust and a grain size of 0.1 $\mu$m. \citet{2010jl} obtained a value of $3.4 \times 10^{16}$ to $1.4 \times 10^{17}$ cm (cf. the dust blackbody radius of SN 2013fc on day $\sim75$, $6 \times 10^{15}$ cm) for the type IIn SN 2010jl assuming graphite grains between 0.001 and 1 $\mu$m in size, respectively, using a bolometric luminosity and temperature similar to those of SN 2013fc ($3 \times 10^{43}$ erg s$^{-1}$ and 10000 K, respectively). All in all, our favoured scenario for the dust emission is that of newly condensing dust in the CDS, as (at least) in the late times of SN 1998S. It is also possible that the formation of new dust contributes to the decline rates during and after the drop from the plateau phase. Dust would condense close to the evaporation temperature, keeping the temperature of the cool component at $\sim2000$ K. Pre-existing dust in the CSM outside the dust evaporation radius may also contribute, but in the \emph{JHK} bands, hot newly-condensed dust probably dominates the emission. In addition to SN 1998S \citep{pozzo, 1998S5}, a late-time NIR excess has been attributed to emission from newly-condensed dust in the CDS at least in the case of the type II-P SN 2007od \citep{07od}. Newly-formed CDS dust has also been suggested as the origin of the NIR excess in the case of the type Ibn SN 2006jc \citep{06jc1, 06jc2, 06jc3} as early as 55 days after explosion. Attenuation of the red wing of the H\,$\alpha$ profile and the steepening of the optical light curve were observed e.g. in two 1988Z-like SNe 2005ip and 2006jd with an early NIR excess and attributed to warm, newly-formed $\sim1500$ K dust by \citet{stritz}. If the dust is condensing in the CDS, it is another sign of CSM interaction, as dust condensation in the ejecta has typically been observed at much later epochs \citep[e.g.][]{dust1, dust2}. Considering the late-time flattening of the NIR light curve, the late-time shrinking of the dust blackbody radius and the constant dust temperature may not be real physical effects but results of the SN still contributing to the \emph{HK} templates. However, this would mean that the NIR emission continued at roughly the same brightness between days $\sim 400$ and $\sim 700$, unlike SN 1998S which declined in all NIR bands at corresponding times.

The similarities between the 1979C/1998S-like SNe described in this paper point to them having similar progenitors as well. The progenitor masses of these events can be estimated by various means. \citet{1979C3} used Padova stellar models to constrain the progenitor of SN 1979C based on the clusters surrounding it, using deep \emph{Hubble Space Telescope} imaging. They arrived at an age estimate of $\sim$ 10 Myr and a progenitor mass of $\sim 17$ -- 18 $M_{\odot}$, consistently with the $\gtrsim 13 M_{\odot}$ suggested by \citet{1979C4} based on mass loss estimates. The He core of the progenitor of SN 1998S was estimated by \citet{1998S3} to be comparable to that of 1987A, which had a $\sim$20 $M_{\odot}$ blue supergiant (BSG) progenitor. The core mass was estimated based on model fitting to CO first-overtone emission. Furthermore, the similarity to II-L SNe may indicate progenitor masses comparable to the 15 -- 18 $M_{\odot}$ suggested for II-L by \citet{smartt2}. \citet{faran2} also suggested $\sim 15 M_{\odot}$ RSG progenitors for type II-L SNe. 

We have also made a very rough assessment of the progenitor mass of SN 2013fc by fitting stellar populations to the location of the SN. The high extinction at the star-forming region at this location gives a reasonable match to an extinction estimated through a comparison with SN 1998S which, together with the unchanging profile of the Na {\sc i} D absorption, implies that SN 2013fc was indeed inside this region. The age estimate of the young population, as in the case of SN 1979C, is $10_{-2}^{+3}$ Myr. We would assume that the progenitor of the SN was the most massive star of this population, if the stars are coeval. This corresponds to an initial mass of $19 \pm 4 M_{\odot}$, assuming solar metallicity and according to the Padova 1994 model tracks \citep{padova94}. Thus the star could have been a massive RSG (or perhaps a BSG) with strong mass loss in the form of a dense wind close to the end of its life. An example of a star such as this would be the red hypergiant VY Canis Majoris \citep[e.g.][]{vycma}. A progenitor mass in this range is roughly consistent with that of the progenitors of SNe 1979C and 1998S, as discussed above. It is also roughly consistent with the upper limit for type II-P/II-L SNe proposed by \citet{smartt2} ($\sim 18 M_{\odot}$).

\section{Conclusions}

We have performed photometric and spectroscopic observations of the 1998S-like supernova SN 2013fc. We have analysed our results and compared the light curves, spectra and possible progenitor scenarios of this event to other similar SNe. 

SN 2013fc is very reminiscent of SN 1998S in terms of its light curve, colour evolution and spectra, and by extension, provides a good match to other 1998S-like SNe such as SN 1979C. The light curve exhibits an initial linear decline followed by a short plateau phase, visible in the \emph{VrRiIJ} bands, and a brightness drop to a tail-like phase where the decline rate is too high for $^{56}$Co decay. The spectra show a nearly featureless blue continuum some three weeks after explosion, followed by the appearance of a strong, asymmetric, broad (FWHM $\sim 8000$ km s$^{-1}$) H\,$\alpha$ emission without a P Cygni profile. We tentatively identify a H\,$\alpha$ absorption around $-15000$ km s$^{-1}$. A strong NIR emission likely originates in hot dust condensing in the CDS that forms at the ejecta-CSM discontinuity. These features indicate interaction between the ejecta and the CSM and a low H envelope mass at the time of explosion, although the presence of early narrow lines cannot be verified, as no pre-maximum spectra of SN 2013fc exist (which is also the case with SN 1979C). SN 2013fc had a peak \emph{B} magnitude of $-20.46 \pm 0.21$; it was 0.9 mag brighter than SN 1998S in the optical, while SN 1979C had a comparable peak brightness. We suggest that a simple $\rho \propto r^{-2}$ density profile of the CSM may be sufficient to explain the observed features of SN 2013fc; this was argued for SN 1998S as well in the latest analyses. If CSM interaction was present, SN 2013fc (like SN 1979C) is likely to be still observable in radio a few years after explosion.

Based on the photometry, we construct a pseudo-bolometric light curve and estimate a total radiated energy of $1.13_{-0.06}^{+0.07} \times 10^{50}$ erg until day 163; $\sim10$ times higher that radiated by normal type II-P SNe, but a factor of at least a few lower than that of bright slowly declining 1988Z-like type IIn SNe. 

SN 2013fc exploded in the circumnuclear ring of a LIRG, behind a high extinction ($A_{V} = 2.9$ mag), coincident with a bright spot in the ring. Stellar population model fits to the coincident source reveal the contribution of a young population of $10_{-2}^{+3}$ Myr, consistent with the progenitor mass ranges proposed for type II-L SNe in general and the proposed progenitor of SN 1979C in particular. Massive RSGs have been suggested as progenitors of 1998S-like SNe; the strong winds of the RSG would then cause the presence of the CSM. We tentatively suggest such a progenitor for SN 2013fc as well.

\section*{Acknowledgements}

We thank the anonymous referee for their comments. 

This work is based on observations collected at the European Organisation for Astronomical Research in the Southern Hemisphere, Chile as part of PESSTO, (the Public ESO Spectroscopic Survey for Transient Objects Survey) ESO program ID 188.D-3003. The observations were made using the ESO New Technology Telescope (NTT) at the La Silla observatory. Some of the observations reported in this paper were obtained with the Southern African Large Telescope (SALT) under the program 2014-1-RSA\_OTH-002. This paper is partly based on observations obtained through the CNTAC proposal CN2013B-68.

This research has made use of the NASA/IPAC Extragalactic Database (NED) which is operated by the Jet Propulsion Laboratory, California Institute of Technology, under contract with the National Aeronautics and Space Administration. We have made use of the Weizmann Interactive Supernova data REPository (WISeREP; http://www.weizmann.ac.il/astrophysics/wiserep/). This research has made use of data products from the Two Micron All Sky Survey, which is a joint project of the University of Massachusetts and the Infrared Processing and Analysis Center/California Institute of Technology, funded by the National Aeronautics and Space Administration and the National Science Foundation.

P.V. acknowledges support from the National Research Foundation of South Africa.

Support for G.P., L.G., C.R.C. and K.T. is provided by the Ministry of Economy, Development, and Tourism's Millennium Science Initiative through grant IC120009, awarded to The Millennium Institute of Astrophysics, MAS.

A.P., N.E.R. and S.B. are partially supported by the PRIN-INAF 2014 with the project ``Transient Universe: unveiling new types of stellar explosions with PESSTO". N.E.R. acknowledges the support from the European Union Seventh Framework Programme (FP7/2007-2013) under grant agreement n. 267251 "Astronomy Fellowships in Italy" (AstroFIt). 

M.F. acknowledges support by the European Union FP7 programme through ERC grant number 320360.

A.G.Y. is supported by the EU/FP7 via ERC grant no. 307260, the Quantum Universe I-Core program by the Israeli Committee for planning and budgeting and the ISF; by Minerva and ISF grants; by the Weizmann-UK ``making connections" program; and by Kimmel and ARCHES awards. 

S.J.S. acknowledges funding from the European Research Council under the European Union's Seventh Framework Programme (FP7/2007-2013)/ERC Grant agreement n$^{\rm o}$ [291222] and STFC grants ST/I001123/1 and ST/L000709/1. R.K. acknowledges support from STFC via ST/L000709/1.

J.V. is supported by Hungarian OTKA Grant NN 107637. T.S. is supported by the OTKA Postdoctoral Fellowship PD112325. 

L.G. acknowledges support by CONICYT through FONDECYT grant 3140566. K.T. was supported by CONICYT through the FONDECYT grant 3150473. C.R.C. acknowledges support from CONICYT through FONDECYT grant 3150238.

J.C.W., G.H.M., J.M.S. and the supernova group at UT Austin are supported by NSF grant AST−1109801. J.M.S. is also supported by an NSF Astronomy and Astrophysics Postdoctoral Fellowship under award AST−1302771.

\section*{Affiliations}

$^{1}$Tuorla Observatory, Department of Physics and Astronomy, University of Turku, V\"{a}is\"{a}l\"{a}ntie 20, FI-21500 Piikki\"{o}, Finland\\
$^{2}$Finnish Centre for Astronomy with ESO (FINCA), University of Turku, V\"{a}is\"{a}l\"{a}ntie 20, FI-21500 Piikki\"{o}, Finland\\
$^{3}$Astrophysics Research Centre, School of Mathematics and Physics, Queen's University Belfast, BT7 1NN, UK\\
$^{4}$Department of Astronomy and The Oskar Klein Centre, AlbaNova University Center, Stockholm University, SE-10691 Sweden\\
$^{5}$South African Astronomical Observatory, PO Box 9, Observatory 7935, Cape Town, South Africa\\
$^{6}$Southern African Large Telescope, PO Box 9, Observatory 7935, Cape Town, South Africa\\
$^{7}$Research School of Astronomy and Astrophysics, Australian National University, Canberra, ACT 2611, Australia\\
$^{8}$ARC Centre of Excellence for All-sky Astrophysics (CAASTRO), Australian National University, Canberra, ACT 2611, Australia\\
$^{9}$Departamento de Ciencias F\'{i}sicas, Universidad Andres Bello, Avda. Rep\'{u}blica 252, Santiago, Chile\\
$^{10}$Millennium Institute of Astrophysics, Chile\\
$^{11}$Department of Physics, University of California, Santa Barbara, Broida Hall, Mail Code 9530, Santa Barbara, CA 93106-9530, USA \\
$^{12}$Las Cumbres Observatory, Global Telescope Network, 6740 Cortona Drive Suite 102, Goleta, CA 93117, USA \\
$^{13}$Department of Optics and Quantum Electronics, University of Szeged, Dom ter 9, Szeged, 6720, Hungary\\
$^{14}$Department of Astronomy, University of Texas at Austin, Austin, TX 78712, USA\\
$^{15}$INAF -- Osservatorio Astronomico di Padova, Vicolo dell'Osservatorio 5, 35122 Padova, Italy\\
$^{16}$Institute of Astronomy (IoA), University of Cambridge, Madingley Road, Cambridge, CB3 0HA United Kingdom\\
$^{17}$Benoziyo Center for Astrophysics, Weizmann Institute of Science, Rehovot 76100, Israel \\
$^{18}$Departamento de Astronom\'{i}a, Universidad de Chile, Camino El Observatorio 1515, Las Condes, Santiago, Chile\\
$^{19}$Nordic Optical Telescope, Apartado 474, 38700 Santa Cruz de La Palma, Spain \\
$^{20}$Department of Astronomy, San Diego State University, San Diego, CA 92182, USA\\
$^{21}$Astrophysics Research Institute, Liverpool John Moores University, IC2, Liverpool Science Park, 146 Brownlow Hill, Liverpool L3 5RF, UK\\
$^{22}$Instituto de Astrof\'{i}sica, Facultad de F\'{i}sica, Pontificia Universidad Cat\'{o}lica de Chile, Casilla 306, Santiago 22, Chile\\
$^{23}$School of Physics and Astronomy, University of Southampton, Southampton SO17 1BJ, UK\\

\label{lastpage}

\end{document}